%% file: cb-vacuum.tex
\documentclass[11pt,a4paper,numbers=noenddot]{article}
\pdfoutput=1

\usepackage[utf8]{inputenc}
\usepackage[T1]{fontenc}
\usepackage{lmodern}

\usepackage[english]{babel}

\usepackage{graphicx}
\usepackage{epsfig}
\usepackage{float}
\usepackage{cite}
\usepackage{amssymb}
\usepackage{amsmath}
\usepackage{amsfonts,amsthm}
\usepackage{url}
\usepackage[dvipsnames]{xcolor}
\usepackage{cancel}
\usepackage{slashed}
\usepackage{nicefrac}
\usepackage[normalem]{ulem}
\usepackage[onehalfspacing]{setspace}
\usepackage{booktabs}
\usepackage{enumitem}
\usepackage{color}

\usepackage[format=plain,labelfont={bf},font={small}]{caption}

\usepackage[colorlinks=true,citecolor=blue,linkcolor=blue]{hyperref}

\newcommand{\email}[1]{\href{mailto:#1}{\nolinkurl{#1}}}

\newcommand{\CP}{\mathrm{CP}}
\newcommand{\CB}{\mathrm{CB}}

\newcommand{\be}{\begin{eqnarray}}
\newcommand{\ee}{\end{eqnarray}}

\newcommand{\ScannerS}{\texttt{ScannerS}}
\newcommand{\BSMPT}{\texttt{BSMPT}}

\newcommand{\gsim}{\raisebox{-0.13cm}{~\shortstack{$>$ \\[-0.07cm]
      $\sim$}}~}
\newcommand{\lsim}{\raisebox{-0.13cm}{~\shortstack{$<$ \\[-0.07cm]
      $\sim$}}~}

\newcommand{\Z}{\mathbb{Z}}
\newcommand{\lr}[1]{ \langle #1 \rangle}

\numberwithin{equation}{section}

\begin{document}

\renewcommand*{\thefootnote}{\fnsymbol{footnote}}

\begin{flushright}
    KANAZAWA-23-08\\
    KA-TP-16-2023
\end{flushright}

\begin{center}
	{\LARGE \bfseries Intermediate Charge-Breaking Phases and Symmetry Non-Restoration in the 2-Higgs-Doublet Model \par}

	\vspace{.7cm}
	Mayumi Aoki\textsuperscript{a,}\footnote{\email{mayumi.aoki@staff.kanazawa-u.ac.jp}},
	Lisa Biermann\textsuperscript{b,}\footnote{\email{lisa.biermann@kit.edu}},
	Christoph Borschensky\textsuperscript{b,}\footnote{\email{christoph.borschensky@kit.edu}},
	Igor P.\ Ivanov\textsuperscript{c,}\footnote{\email{ivanov@mail.sysu.edu.cn}},
	Margarete M\"uhlleitner\textsuperscript{b,}\footnote{\email{margarete.muehlleitner@kit.edu}},
	Hiroto Shibuya\textsuperscript{a,}\footnote{\email{h.shibuya.phys@gmail.com}}

	\vspace{.3cm}
	\textit{
		\textsuperscript{a }Institute for Theoretical Physics, Kanazawa University, Kanazawa 920-1192, Japan\\[.2em]
		\textsuperscript{b }Institute for Theoretical Physics, Karlsruhe Institute of Technology, Wolfgang-Gaede-Str.\ 1, 76131 Karlsruhe, Germany\\[.2em]
		\textsuperscript{c }School of Physics and Astronomy, Sun Yat-sen University, 519082 Zhuhai, China
	}
\end{center}

\renewcommand*{\thefootnote}{\arabic{footnote}}
\setcounter{footnote}{0}

\vspace*{0.3cm}
\begin{abstract}
The Higgs potentials of extended Higgs sectors exhibit a complex and interesting vacuum structure. When travelling back in time, i.e.\ going to higher temperatures, the structure may change and exhibit interesting phase patterns and sequences of phases related to the respective minima of the potential. The investigation of the vacuum structure can give us indirect insights in beyond-Standard-Model physics and the evolution of the Universe. In this paper, we investigate the possibility of an intermediate charge-breaking (CB) phase in the 2-Higgs-Doublet Model (2HDM) type I. The existence has been reported previously by using a simple potential setup. We here confirm that the intermediate CB phase can still exist when using the one-loop corrected effective potential including thermal masses. We discuss its features and the relation with $SU(2)$ symmetry (non-)restoration as well as its consistency with the current experimental data. Lastly, we 
show for some selected benchmark points the rich and interesting phase patterns and sequences that the 2HDM can undergo during its evolution from the early Universe to today's electroweak vacuum. 
\end{abstract}

\thispagestyle{empty}
\vfill
\pagebreak

\tableofcontents

\input{1-introduction.tex}
\input{2-2hdm.tex}
\input{3-chargebreaking.tex}
\input{4-symrestoration.tex}
\input{5-numerics.tex}
\input{6-conclusions.tex}

\appendix
\input{7-appendix.tex}

\bibliographystyle{spphys_custom}
\bibliography{bibliography}

\end{document}

%% file: 1-introduction.tex
\section{Introduction}\label{s:intro}

How did the hot early Universe evolve around the electroweak epoch? 
The answer to this question is critically important not only to our understanding of the present-day Universe 
but also to the quest for physics beyond the Standard Model (SM). In particular, this evolution is a remarkable
testbed, complementary to collider searches, 
for the models beyond the SM which make use of extended scalar sectors.
Around the electroweak epoch, the Universe may have experienced a sequence of phase transitions 
of different nature \cite{Guo:2020grp,Athron:2023xlk},
possibly evolving through exotic intermediate phases.
Violent phase transitions produced primordial gravitational waves (GW), 
which may be within the reach of future space-borne GW observatories.
A strong first-order phase transition within a CP-violating Higgs sector is a key part
of electroweak baryogenesis \cite{Kuzmin:1985mm,Cohen:1990it,Cohen:1993nk,Quiros:1994dr,Rubakov:1996vz,Funakubo:1996dw,Trodden:1998ym,Bernreuther:2002uj,Morrissey:2012db}, in which the matter-antimatter asymmetry was seeded through 
the true vacuum bubble wall sweeping the hot plasma.
Phase transitions induced by the vacuum structure of the finite-temperature scalar potential
could also be connected with symmetries, which in turn could stabilize scalar dark matter candidates.

All these interrelated processes dramatically depend on the structure of the Higgs sector of our Universe. 
The SM is unable to correctly account for the above astroparticle and cosmological phenomena \cite{Kajantie:1996mn,Csikor:1998eu}, 
but minimally extended Higgs sectors can.
For example, within the 2-Higgs-Doublet Model (2HDM) \cite{Lee:1973iz,Branco:2011iw,Ivanov:2017dad},
one can achieve a strong first-order electroweak (EW) phase transition \cite{Dorsch:2013wja,Dorsch:2014qja,Basler:2016obg,Dorsch:2017nza,Basler:2017uxn,Andersen:2017ika,Bernon:2017jgv,Basler:2019iuu,Su:2020pjw,Fabian:2020hny,Basler:2021kgq,Enomoto:2021dkl,Kanemura:2022ozv,Atkinson:2022pcn,Song:2022xts,Anisha:2022hgv,Enomoto:2022rrl,Biekotter:2022kgf,Kanemura:2023juv,Astros:2023gda,Goncalves:2023svb} 
and electroweak baryogenesis, provided there is sufficient CP violation in the scalar sector.

Extended Higgs models also allow for multi-step phase transitions in the early Universe.
This possibility that was already described by Weinberg in \cite{Weinberg:1974hy}, was later analyzed,
within specific models, in \cite{Ginzburg:2009dp, Patel:2012pi, Blinov:2015sna,Morais:2019fnm, Aoki:2021oez, Benincasa:2022elt, Shibuya:2022xkj,Liu:2023sey,Aoki:2023xnn}.
A sequence of phase transitions could lead to the remarkable opportunity of a charge-breaking (CB) phase at intermediate temperatures. 
In this phase, the vacuum expectation values (VEVs) of the Higgs fields
break the electroweak symmetry completely, the photon becomes massive, and the electric charges of the fermions
are no longer conserved. This exotic phase does not correspond to the present-day zero-temperature vacuum, which must be electrically neutral,
but it may have been an important episode of the thermal history of the Universe.

An additional motivation for investigating CB phases comes from the puzzling origin of the cosmological magnetic fields 
(for more details, see reviews, e.g., Refs.~\cite{Widrow:2002ud, Durrer:2013pga, Subramanian:2015lua, Vachaspati:2020blt}).
Various scenarios have been proposed, including the possibility that magnetic fields were seeded by the magneto-hydrodynamic turbulence 
from the first-order EW phase transition, see e.g.\ \cite{Kahniashvili:2009mf, Caprini:2009yp, Binetruy:2012ze}.
However, an intermediate CB phase, an electromagnetically disruptive event, could have played a role in this process.

The 2HDM is the simplest Higgs sector in which the CB phase can take place at finite temperature.
In fact, the 2HDM has a very rich vacuum structure, with the possibility of charge- or CP-breaking vacua and even normal vacua (i.e. vacua conserving charge and CP) that could have a VEV different from 246~GeV. However, at vanishing temperature it has been shown that whenever a normal vacuum exists, any charge or CP-breaking stationary point that possibly exists, must be a saddle point and lie above the normal minimum \cite{Ferreira:2004yd,Barroso:2005sm}. There exists though the possibility that two normal vacua coexist with one another \cite{Ivanov:2006yq,Barroso:2007rr,Ivanov:2007de}. In case our vacuum is not the global minimum, i.e.~is a so-called panic vacuum, the Universe would be in a metastable state and could tunnel to the deeper normal vacuum with a VEV different from 246~GeV. In \cite{Barroso:2012mj,Barroso:2013awa}, the authors presented the conditions that the parameters of the potential need to obey to avoid the presence of a panic vacuum in the softly-broken version of the 2HDM. They also showed that for this model the current LHC data already constrain the model to necessarily be in the global minimum of the theory. 
At non-zero temperature, however, the vacuum structure can change considerably, and the features derived for $T=0$ do not hold anymore.
The conditions and the gross features of the CB phase were analyzed in \cite{Ivanov:2008er,Ginzburg:2009dp}
within the high-temperature approximation of the 2HDM potential, in which the tree-level potential is retained
and only the quadratic mass parameters $m_{ii}^2$ ($i=1,2$) receive corrections from non-zero temperature $T$, 
$m_{ii}^2 \to m_{ii}^2(T) = m_{ii}^2 + c_i T^2$.
These works found that, within the adopted approximation, 
the following sequence of thermal phase transitions occurs within a certain region of the 2HDM potential parameters, ending in the EW vacuum with today's vacuum expectation value of 246~GeV at $T=0$:
\begin{equation}
\mbox{EW symmetric (high $T$)} \to \mbox{neutral} \to \mbox{charge-breaking}\to \mbox{EW vacuum ($T=0$)} \,.
\label{PT-sequence}
\end{equation}
The phase transitions leading from the broken neutral phase into and out of the CB phase were found to be of second order.

It turns out that this tantalizing possibility has not yet been properly reassessed
with the state-of-the-art numerical codes, which examine the thermal evolution of the loop-corrected
effective potential of the 2HDM, and with all the LHC data and constraints we now have. 
This is what we undertake in the present work. 
In particular, we will address the following questions:
\begin{itemize}
\item 
Do intermediate CB phase scenarios indeed exist within the 2HDM? If so, in which parameter regions?
\item
Do the 2HDM versions with an intermediate CB phase comply with the collider constraints on the Higgs sector?
Do they lead to any characteristic collider predictions?
\item
What sequences of thermal phase transitions do these models exhibit? 
In particular, do they always lead to the electroweak symmetry restoration at high $T$?
\end{itemize}
In this paper, we investigate these issues within the CP-conserving 2HDM, treated with the finite temperature loop-corrected effective potential level including thermal masses within 
the properly adapted \texttt{BSMPT} code~\cite{Basler:2018cwe,Basler:2020nrq}.
Guided by the phase diagram technique developed in the tree-level studies \cite{Ivanov:2008er,Ginzburg:2009dp}, 
we will search for parameter space regions which exhibit an intermediate CB phase. 
We will then analyze the main features of such models and confront them with the LHC Higgs results. 
On the way, we will also discuss the relation between the CB phase and $SU(2)$ symmetry non-restoration at high temperature. 
Note however, that in this paper we will not discuss the intriguing gravitational waves and electroweak baryogenesis opportunities offered by the charge-breaking phase nor the implications that the sequence of found exotic vacuum phases may have on the cosmological evolution and dark matter generation. These interesting and wide ranging implications are delegated to dedicated future work.

The contents of this paper are as follows. In Sec.~\ref{s:cpc2hdm}, we remind the reader of the CP-conserving 2HDM with a softly broken $\Z_2$ symmetry. In Sec.~\ref{s:chargebreaking}, we discuss the essence of the phenomenon in a toy model (Type I 2HDM with an exact $\Z_2$ symmetry),
which provides us with a clear qualitative intuition of how to search for the CB phase.
Then, in Sec.~\ref{s:symrestoration}, we proceed with the thermal effective potential formalism and its relation with symmetry non-restoration. 
In Sec.~\ref{s:numerics}, we present numerical results from a parameter search, show the temperature evolution of several benchmark models,
and discuss their collider implications. Finally, we draw our  conclusions in Sec.~\ref{s:conclusion}.

%% file: 2-2hdm.tex
\section{The CP-conserving 2HDM}\label{s:cpc2hdm}
In our study, we will consider the CP-conserving 2-Higgs Doublet Model (2HDM) with a softly broken $\Z_2$ symmetry \cite{Lee:1973iz} (see e.g.\ \cite{Branco:2011iw} for a review). Its tree-level potential is given by
\begin{equation}
	\begin{split}
		V_{\text{tree}} &= m_{11}^2\Phi_1^\dag \Phi_1 + m_{22}^2\Phi_2^\dag \Phi_2 - m_{12}^2\left(\Phi_1^\dagger\Phi_2 + \mathrm{h.c.}\right) + \frac{\lambda_1}{2}\left(\Phi_1^\dag \Phi_1\right)^2 + \frac{\lambda_2}{2}\left(\Phi_2^\dag \Phi_2\right)^2\\
		&\quad + \lambda_3\left(\Phi_1^\dag \Phi_1\right)\left(\Phi_2^\dag \Phi_2\right) + \lambda_4\left(\Phi_1^\dag \Phi_2\right)\left(\Phi_2^\dag \Phi_1\right) + \frac{\lambda_5}{2}\left[\left(\Phi_1^\dag \Phi_2\right)^2 + \mathrm{h.c.}\right],
  \end{split}\label{eq:v2hdm}
\end{equation}
where all parameters are assumed to be real. After electroweak symmetry breaking (EWSB) the two $SU(2)$ scalar doublets $\Phi_1$ and $\Phi_2$ obtain VEVs $\bar{\omega}_j$ ($j=$~1, 2, CP, CB) about which the Higgs fields can be expanded in terms of the fields $\rho_i$, $\eta_i$, $\zeta_i$, and $\psi_i$ ($i=1,2$), 
\begin{align}
	\Phi_1 &= \frac{1}{\sqrt{2}}
		\begin{pmatrix}
			\rho_1 + i\eta_1 \\
			\zeta_1 + \bar\omega_1 + i\psi_1
		\end{pmatrix},\label{eq:phi1}\\
	\Phi_2 &= \frac{1}{\sqrt{2}}
		\begin{pmatrix}
			\rho_2 + \bar\omega_\CB + i\eta_2 \\
			\zeta_2 + \bar\omega_2 + i\left(\psi_2 + \bar\omega_\CP\right)
		\end{pmatrix}.\label{eq:phi2}
\end{align}
Without loss of generality we rotated the CP-violating part $\bar{\omega}_{\text{CP}}$ of the VEVs to the second doublet exclusively. Adopting the most general approach, we also introduced a charge-breaking VEV $\bar\omega_{\text{CB}}$, which 
without loss of generality can be taken real \cite{Ginzburg:2009dp}. Our present vacuum at zero temperature $T = 0$ is denoted as
\begin{eqnarray}
v_j = \bar\omega_j|_{T=0}
\end{eqnarray}
and given by
\begin{equation}\label{eq:vevpresent}
	v_\CB = v_\CP = 0\quad\text{and}\quad v \equiv \sqrt{v_1^2 + v_2^2} = 246.22\text{ GeV},
\end{equation}
with the ratio of the EW VEVs defining
\begin{equation}\label{eq:tanbeta}
\tan \beta \equiv \frac{v_2}{v_1} \;.
\end{equation}
We assume CP conservation, and the CB VEV has to be zero in a charge-conserving vacuum.
At $T=0$, after rotating to the mass eigenstates, we hence recover the usual CP-even ($h,H$), CP-odd ($A$) and charged Higgs ($H^\pm$) fields of the CP-conserving 2HDM.

In the $SU(2)$ symmetric phase, all VEVs are zero, $\bar\omega_i = 0$.
The necessary and sufficient conditions for the tree-level potential of Eq.~\eqref{eq:v2hdm} to be bounded from below (BFB) are well known and given by \cite{Deshpande:1977rw,Klimenko:1984qx},
\begin{equation}
\lambda_{1} > 0\,, \quad \lambda_{2} > 0\,, \quad \sqrt{\lambda_1\lambda_2} + \lambda_3 > 0\,, \quad 
\sqrt{\lambda_1\lambda_2} + \lambda_3 + \lambda_4 - |\lambda_5|> 0\,.\label{BFB}
\end{equation}
In order to allow for a perturbative treatment of the interactions, the model must not violate unitarity and thus not be strongly coupled. Constraints on  the scalar couplings can be derived from limits on the eigenvalues of the $2\to 2$ scattering matrix \cite{Lee:1977eg,Kanemura:1993hm,Akeroyd:2000wc,Ginzburg:2003fe}. 

\begin{table}[tp]
    \centering
    \begin{tabular}{rccc}
        \toprule
        & $u$-type & $d$-type & leptons \\
        \midrule
        Type I & $\Phi_2$ & $\Phi_2$ & $\Phi_2$ \\
        Type II & $\Phi_2$ & $\Phi_1$ & $\Phi_1$ \\
        Lepton-Specific & $\Phi_2$ & $\Phi_2$ & $\Phi_1$ \\
        Flipped & $\Phi_2$ & $\Phi_1$ & $\Phi_2$ \\
        \bottomrule
    \end{tabular}
    \caption{The four types for the Yukawa couplings in the 2HDM with a softly broken $\Z_2$ symmetry. The last three columns indicate the Higgs doublet to which the respective fermions couple.}
    \label{tab:yuktypes}
\end{table}
The interactions between the Higgs doublets and the fermion fields are given by the Yukawa couplings. To avoid the appearance of flavour-changing neutral currents (FCNC) at tree level, we extend the $\Z_2$ symmetry also to the fermion sector and each of the three classes of fermions, i.e.\ up-type quarks, down-type quarks, and leptons, is restricted to couple to one of the two Higgs doublets only. Depending on the assignment of the $\Z_2$ charges, there are four different types of Yukawa couplings in the 2HDM, as listed in Table~\ref{tab:yuktypes}.

In the following, we adopt the 2HDM type I. In the context of our work of phase transitions, the differences between the types of Yukawa couplings are largely irrelevant, as in all types the Yukawa coupling of the top quark, which has the strongest impact, is the same, and the influence from the Yukawa couplings of the other fermions is negligible. The chosen type of the 2HDM has an impact, however, on how much the model is constrained, and  thereby on how much of the parameter space is still  allowed, cf.\ e.g.\ \cite{Abouabid:2021yvw}.

%% file: 3-chargebreaking.tex
\section{Intermediate charge-breaking phase}\label{s:chargebreaking}

\subsection{A toy model: the phase diagram}\label{s:chargebreaking:phasediag}

Before we undertake a full analysis, 
it is instructive to gain an intuitive understanding of the phenomenon  
within a toy model.
This model is given by the scalar sector of the 2HDM with an exact $\Z_2$ symmetry, 
which we treat at tree level
and in which, at finite temperature, we include only the $T^2$ terms 
to the quadratic parameters of the potential.
This toy model can accommodate the intermediate CB phase, 
which can be tracked analytically
and easily visualized as a trajectory on a two-dimensional phase diagram.
It matches, in part, the picture described in \cite{Ivanov:2008er,Ginzburg:2009dp,Ginzburg:2010wa}. 
But, in contrast to these papers, we use it here only as a starting point 
before moving to the full effective potential treatment of the finite temperature 2HDM.

The 2HDM with the exact $\Z_2$ symmetry is given by the scalar potential of Eq.~\eqref{eq:v2hdm} with the soft-breaking parameter set to zero, $m_{12}^2 = 0$. 
Within the toy model, we assume that the potential has this functional form at any $T$,
with only $m_{11}^2$ and $m_{22}^2$ acquiring the finite temperature corrections. 

To minimize this potential, the usual procedure would be to assume that the minimum is neutral, 
to parametrize it in terms of the VEVs of Eq.~\eqref{eq:vevpresent}, and to express the quadratic
parameters via $v$.
However, in this study we look into the richer spectrum of options available at intermediate temperatures,
including the charge-breaking phase.
This is why we do not make any assumption about the VEVs and just proceed with the direct minimization of this potential.
Depending on the values of the quadratic terms, we can have global minima of different kinds, corresponding to 
the EW symmetric ($\lr{\Phi_1} = \lr{\Phi_2} = 0$), the neutral, the CP- or the charge-breaking vacuum. In the following, we want to concentrate on the possibility of a CB phase so that we do not consider a CP-breaking phase here further.
We will present the EW-symmetric, the neutral, and the CB regions 
in the phase diagram of the model, that is, in  
the two-dimensional plane $(m_{11}^2, m_{22}^2)$ for a specific set of the quartic coefficients $\lambda_i$.
Since only the quadratic parameters evolve with the temperature,
we can represent the thermal evolution of the toy model as a trajectory in this plane
ending with the present-day $T=0$ point. 

The analytical computation is straightforward and yields the necessary and sufficient conditions for the toy model potential
to admit a charge-breaking minimum, see also \cite{Ivanov:2008er} for the analysis of the most general 2HDM.
There are two groups of conditions.
First, in order for the CB minimum to appear, the quartic coefficients must satisfy, in addition to the BFB constraints \eqref{BFB}, 
the following inequalities:
\begin{equation}
\sqrt{\lambda_1\lambda_2} - \lambda_3 > 0\,, \quad \lambda_4 > |\lambda_5|\,.\label{CB-condition-1}
\end{equation}
In particular, we get $|\lambda_3| < \sqrt{\lambda_1\lambda_2}$, but the sign of $\lambda_3$ is not fixed.
Second, if these conditions are satisfied, the quadratic parameters must comply with the following constraints:
\begin{eqnarray}
m_{11}^2 \sqrt{\lambda_2} + m_{22}^2 \sqrt{\lambda_1} < 0\,, \quad
m_{11}^2 < m_{22}^2\frac{\lambda_3}{\lambda_2}\,, 
\quad m_{22}^2 < m_{11}^2\frac{\lambda_3}{\lambda_1}\,.\label{CB-condition-2} 
\end{eqnarray}
Notice that the signs of $\lambda_3$, $m_{11}^2$, and $m_{22}^2$
are not fixed {\em a priori}, so that both sign choices of $\lambda_3$ should be explored.
\begin{figure}[tp]
	\centering
	\includegraphics[width=0.45\textwidth]{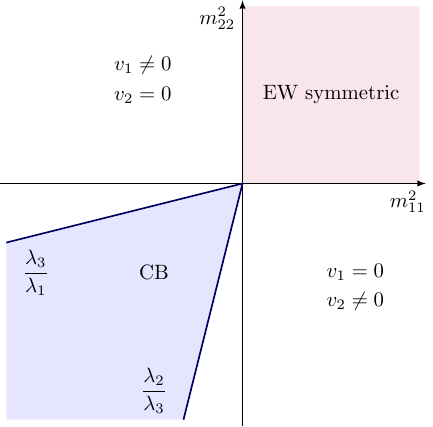}
	\qquad
	\includegraphics[width=0.45\textwidth]{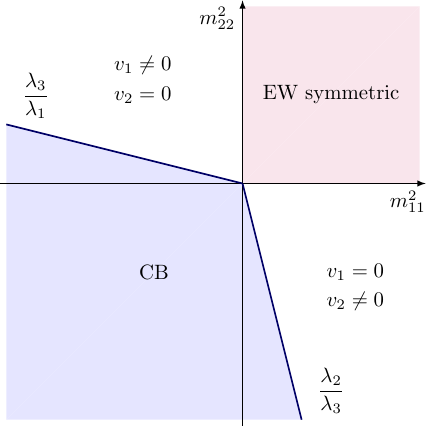}
	\caption{EW symmetric vacuum (red) and CB vacuum (blue) regions in the plane $(m_{11}^2, m_{22}^2)$ for $\lambda_3 > 0$ (left)
	and  $\lambda_3 < 0$ (right). Labels at the lines should be read as $m_{22}^2 = m_{11}^2 \,\times$~the label.}
	\label{fig-regions-1}
\end{figure}

The resulting phase diagrams are shown in Fig.~\ref{fig-regions-1}
for $\lambda_3 > 0$ (left) and $\lambda_3 < 0$ (right). 
The necessary and sufficient conditions for the EW symmetric vacuum (red regions) are trivial: $m_{11}^2 > 0$ and $m_{22}^2 > 0$.
If at least one of these coefficients is negative, we can attain a CB minimum (blue wedge regions),
provided the inequalities \eqref{CB-condition-2} are fulfilled.
We stress once again that these phase diagrams apply only to the case when the quartic coefficients 
satisfy \eqref{CB-condition-1}; should this not be the case, no CB vacuum is possible.
It is known since \cite{Ferreira:2004yd,Barroso:2005sm,Ivanov:2006yq} that a charge-breaking minimum, if present, is the unique minimum of the potential. 
Thus, within the 2HDM, at least at tree level, we never encounter the situation when a charge-breaking and a neutral minimum
would be simultaneously present.

The blank regions on both plots correspond to a neutral vacuum. Within the toy model, 
this vacuum corresponds either to $v_1 = 0$ or $v_2 = 0$.
A stationary point with both $v_1, v_2 \not = 0$ also exists, but it is always a saddle point 
due to the negative charged Higgs mass squared.
The region above the blue CB wedge corresponds to the global minimum $v_1 \not = 0$, $v_2 = 0$.
Since in the 2HDM type I all the fermions get their masses only from the second doublet, 
this option is unphysical.
The other blank region, to the right of the CB blue wedge, corresponds to the global minimum $v_1 = 0$, $v_2 \not= 0$.
It is here that the zero-temperature situation must reside.
Within this region, the VEV, the SM-like Higgs mass, and the charged Higgs mass are 
\begin{equation}
v^2 = v_2^2 = \frac{2|m_{22}^2|}{\lambda_2}\,, \quad \frac{m_{h_{\text{SM}}}^2}{v^2} = \lambda_2\,,
\quad
\frac{m_{H^\pm}^2}{v^2} = \frac{1}{2}\lambda_3 \left(1 - \frac{m_{11}^2}{m_{22}^2}\frac{\lambda_2}{\lambda_3}\right)\,.
\label{blank-region-2}
\end{equation}
Taking the experimentally known values of $m_{h_{\text{SM}}} =125.09$~GeV \cite{ParticleDataGroup:2022pth} and $v$, we get $\lambda_2 \approx 1/4$.
Notice also that if one approaches the boundary separating the neutral and CB vacuua, 
$m_{22}^2/m_{11}^2 = \lambda_2/\lambda_3$, the charged Higgs mass vanishes. 

This blank region corresponds to the version of the 2HDM known as the inert doublet model (IDM) \cite{Deshpande:1977rw,Ma:2006km,Barbieri:2006dq,LopezHonorez:2006gr}.
It possesses a scalar dark matter candidate and has been extensively studied. 
However in the particular version of the IDM with $\lambda_4 - |\lambda_5| > 0$, the one with which we deal here, 
it is the charged Higgs boson that becomes the lightest scalar from the inert doublet and plays the role
of the dark matter candidate. 
This is why the IDM was considered in  \cite{Ginzburg:2010wa} incompatible with the intermediate charge breaking vacuum phase.
In the present paper, we take this scenario only as a toy model before moving on to the softly broken $\Z_2$ symmetry.

\subsection{A toy model: temperature evolution}\label{s:chargebreaking:tempevo}

Within the toy model, we take the following one-loop high-$T$ corrections to the quadratic terms, 
\begin{equation}\label{eq:highTthermalparams}
m_{11}^2(T) = m_{11}^2 + c_1T^2\,, \quad
m_{22}^2(T) = m_{22}^2 + c_2T^2\,, 
\end{equation}
and adopt them for all temperatures.
Within the 2HDM type I, the coefficients $c_1$ and $c_2$ have the following form:
\begin{align}
    c_1 &= \frac{1}{12}\left(3\lambda_1 + 2\lambda_3 + \lambda_4\right) + \frac{1}{16}\left(3g^2 + g'^2\right)\,,\label{eq:c1coeff}\\
    c_2 &= \frac{1}{12}\left(3\lambda_2 + 2\lambda_3 + \lambda_4\right) + \frac{1}{16}\left(3g^2 + g'^2\right) + \frac{1}{12}\left(y_\tau^2 + 3y_b^2 + 3y_t^2\right)\,,\label{eq:c2coeff}
\end{align}
where $g'$ and $g$ denote the $U(1)_Y$ and $SU(2)_L$ gauge couplings, respectively, and $y_x$ $(x=\tau,b,t$) the Yukawa couplings to the $\tau$ lepton, the bottom and the top quark, respectively.  
Note that the scalar self-interaction contributions to $c_1$ and $c_2$ differ only in $\lambda_1$ vs. $\lambda_2$,
the gauge boson contributions are identical, and the Yukawa couplings of $\tau$, $b$, $t$ affect only $m_{22}^2(T)$.

In order to arrange for the phase transition sequence given in Eq.~\eqref{PT-sequence}, which we now track in the reverse order, 
from $T=0$ to high temperatures, 
we must start at a point $(m_{11}^2, m_{22}^2)$ lying within the blank region below the wedge.
As $T$ grows, the point $(m_{11}^2(T), m_{22}^2(T))$ will follow a straight ray, which must enter and then exit 
the charge-breaking wedge (the blue region).
If we insist on high-$T$ restoration of the EW symmetry, the ray must eventually enter the red quadrant. 

It follows directly from the phase diagrams Fig.~\ref{fig-regions-1} that this sequence is impossible for $\lambda_3 < 0$
because, in this case,
a straight ray starting from anywhere in the blank region and crossing the blue wedge 
will unavoidably continue in the blank region, leading to the remarkable phenomenon 
of non-restoration of the EW symmetry at high temperatures. 
The scenario of symmetry non-restoration has been discussed in the literature,
starting from example 3 of the seminal paper \cite{Weinberg:1974hy} up to more recent studies such as \cite{Espinosa:2004pn,Ramsey-Musolf:2017tgh,Meade:2018saz,Baldes:2018nel,Glioti:2018roy,Matsedonskyi:2020mlz,Carena:2021onl,Biekotter:2021ysx,Biekotter:2022kgf}.
Although it does not seem to be outright excluded by experiment,
let us insist in this section on EW symmetry restoration at high temperatures.
Then, we must take a positive $\lambda_3$ and consider Fig.~\ref{fig-regions-1} (left).
The positive $\lambda_3$ leads to $c_1 > 0$, $c_2 > 0$, 
so that EW symmetry restoration is not only possible but also guaranteed.

Next, in order to actually cross the blue wedge region, one needs to select the starting zero-temperature point 
in the blank region below the wedge and make sure that the ray rises steeper than the wedge:
\begin{equation}
\frac{c_2}{c_1} > \frac{|m_{22}^2|}{|m_{11}^2|} > \frac{\lambda_2}{\lambda_3}.
\label{toy-model-condition}
\end{equation}
Satisfying these inequalities using only bosonic contributions and, at the same time, avoiding a dangerously small $m_{H^\pm}^2$ 
is not an easy task. Indeed, let us first check the case of $\lambda_1 = \lambda_2$ and neglect the fermion contributions.
Then the inequality fails: 
$c_2/c_1 = 1$, while $\lambda_2/\lambda_3 \ge 1$ because $\lambda_3$ is limited from above  
by $\lambda_{3,\,\text{max}} = \sqrt{\lambda_1\lambda_2}$, see Eq.~\eqref{CB-condition-1}.
To satisfy the inequality, we need to push $c_2/c_1$ above $\lambda_2/\lambda_{3,\,\text{max}} = \sqrt{\lambda_2/\lambda_1}$.
This can be done by increasing $\lambda_1$ (to suppress $\lambda_2/\lambda_{3,\,\text{max}}$) 
and, at the same time, to keep $\lambda_4$ large to prevent $c_2/c_1$ from becoming too low. 
A large $\lambda_1$ also allows us to increase $\lambda_{3,\,\text{max}}$ and, therefore, $\lambda_3$ 
in order to avoid a too light charged Higgs.
Also, choosing $\lambda_{3}$ close to $\lambda_{3,\,\text{max}}$ 
makes the CB blue wedge narrow 
and allows the trajectory to cross it even if the zero-temperature starting point lies 
significantly below the blue wedge.

\begin{figure}[tp]
	\centering
	\includegraphics[width=0.6\textwidth]{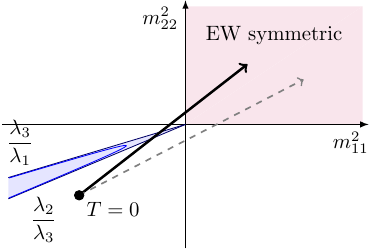}
	\caption{A possible trajectory (solid thick line) within the toy model exhibiting the charge breaking phase at intermediate temperatures;
	the numerical values are given in Eq.~\eqref{example-trajectory}.
	If one drops the top-quark contribution to $c_2$, one will miss the charge-breaking region (dashed line).
        The blue hyperbola inside the wedge shows the CB contour after the inclusion of the soft breaking $m_{12}^2$ term.}
	\label{fig-regions-2}
\end{figure}

Figure~\ref{fig-regions-2} shows a possible trajectory which satisfies these constraints.
Here, we took 
\begin{equation}
\lambda_1 = 2,\quad \lambda_2 = 0.25, \quad \lambda_3 = 0.6, \quad \lambda_4 = 2.8, \quad
\frac{|m_{22}^2|}{|m_{11}^2|} = \frac{2}{3}\,.
\label{example-trajectory}
\end{equation}
With $c_2/c_1 \approx 0.78$ and $\lambda_2/\lambda_3 \approx 0.42$, all the inequalities \eqref{toy-model-condition} hold. 
The charged Higgs mass is then $m_{H^\pm} \approx 82$ GeV.
Trying to significantly increase $m_{H^\pm}$ would force us either to push $\lambda_3$ to even larger values, 
which would unavoidably require a larger $\lambda_1$ and, as a result, a less steep trajectory,
or to select a larger value of $|m_{22}^2|/|m_{11}^2|$ for the starting point,
which will again put the intermediate CB phase at risk.

Notice also the instrumental role of the top-quark contribution to $c_2$. Without it,
one would observe a less steep trajectory, shown by the dashed line in Fig.~\ref{fig-regions-2}, 
with $c_2/c_1 \approx 0.5$, and one would fail to cross the CB wedge region.

Let us now summarize several lessons we have learned from the toy model.
\begin{itemize}
\item
If we insist on EW symmetry restoration at high temperatures, we need $\lambda_3 > 0$.
\item 
The preferred choice of the free parameters, which leads to an intermediate CB phase, 
is $\lambda_1 \gg \lambda_2 \approx 1/4$, $\lambda_3$ close to $\sqrt{\lambda_1\lambda_2}$,
and large $\lambda_4$.
\item
The zero-temperature point should correspond to $|m_{22}^2|/|m_{11}^2|$ not too close to the
CB/neu\-tral boundary in order to avoid a dangerously small charged Higgs mass.
At the same time, $|m_{22}^2|/|m_{11}^2|$ cannot be too large in order not to miss the intermediate CB phase altogether.
\item
The top-quark contribution to the thermal evolution coefficient plays a key role in achieving 
an intermediate CB phase. If we dropped it from the thermal corrections, 
we would be forced either to dramatically increase $\lambda_4$ or to choose an unacceptably small charged Higgs mass
at the starting point.
\end{itemize}

\subsection{Softly broken $\Z_2$ symmetry}\label{s:chargebreaking:softz2hdm}

The full model we consider in this work, which differs from the toy model by the presence of the soft $\Z_2$ breaking $m_{12}^2$ term,
can also be analyzed within the same geometric framework, at least at tree level (an extension of the geometric approach to the one-loop effective potential was recently developed in \cite{Cao:2022rgh,Cao:2023kgq}, 
but we do not rely on it in this work).
The phase diagram now acquires two additional dimensions defined by the real and imaginary parts of $m_{12}^2$.
This tree-level analysis in its full generality was conducted in \cite{Ivanov:2008er}, 
and criteria for the charge breaking vacuum were established. 
Here, we repeat its main conclusions for the real $m_{12}^2$ case.
\begin{itemize}
\item 
All the constraints on the quartic coefficients, including the BFB conditions and the necessary CB conditions Eq.~\eqref{CB-condition-1},
remain unchanged.
\item
The first condition from Eq.~\eqref{CB-condition-2} is also intact: $m_{11}^2 \sqrt{\lambda_2} + m_{22}^2 \sqrt{\lambda_1} < 0$.
However, the region with the CB vacuum now corresponds to the cone in the $(m_{11}^2,m_{22}^2,m_{12}^2)$ space defined by
\begin{equation}\label{CB-condition-3}
	\quad \frac{\mu_1^2}{a_1^2} + \frac{\mu_2^2}{a_2^2} < 1\,,
\end{equation}
where 
\begin{eqnarray}
		&&\mu_1 = \left|\frac{2\sqrt[4]{\lambda_1\lambda_2} m_{12}^2}{m_{11}^2\sqrt{\lambda_2} + m_{22}^2\sqrt{\lambda_1}}\right|,
		\quad a_1 = \frac{\lambda_4 + \lambda_5}{\sqrt{\lambda_1\lambda_2} + \lambda_3}\,,\nonumber\\[2mm]
		&&\mu_2 = \left|\frac{m_{11}^2\sqrt{\lambda_2} - m_{22}^2\sqrt{\lambda_1}}{m_{11}^2\sqrt{\lambda_2} + m_{22}^2\sqrt{\lambda_1}}\right|,
		\quad a_2 = \frac{\sqrt{\lambda_1\lambda_2} - \lambda_3}{\sqrt{\lambda_1\lambda_2} + \lambda_3}\,.
\label{CB-condition-4}
\end{eqnarray}
Notice that in the case $m_{12}^2 = 0$ we recover the conditions \eqref{CB-condition-2}.
\end{itemize}
As in the toy model, one can adopt the temperature corrections as in Eq.~\eqref{eq:highTthermalparams} so that $m_{11}^2$
and $m_{22}^2$ evolve with $T$, while the soft breaking coefficient $m_{12}^2$ is unaffected by finite temperature effects in the high-$T$ approximation. 
As a result, when tracking the temperature evolution, we still explore the two-dimensional phase diagram 
$(m_{11}^2,m_{22}^2)$ taken at a fixed $m_{12}^2$.
However, instead of the CB wedge, we now have the hyperbolic conical section illustrated by the blue curve in Fig.~\ref{fig-regions-2}.
The larger $m_{12}^2$, the further the CB region retracts from the origin. There is a single blank region now which still corresponds to neutral vacua but with both $v_1\not = 0$ and $v_2\not = 0$. 
The trajectory, which still begins with a $T=0$ point in the blank region below the hyperbola, 
still needs to cross the CB region.
Clearly, in order not to miss it, the value of $m_{12}^2$ should not be too large.

%% file: 4-symrestoration.tex
\section{Effective potential and symmetry non-restoration}\label{s:symrestoration}
While so far, we only discussed the high-temperature behaviour of the tree-level potential by introducing $T$-dependent quadratic parameters $m_{11}^2$ and $m_{22}^2$, in the next step, we want to study the behaviour of intermediate CB phases for the one-loop-corrected effective potential, which will be introduced in the following. We furthermore discuss briefly the (non-)restoration of the EW symmetry at high temperatures for the one-loop effective potential and derive conditions to distinguish between the possibilities of restoration and non-restoration, so that in our numerical analysis in Sec.~\ref{s:numerics}, we can study possible connections between finding viable points with intermediate CB phases and (non-)restoration of the EW symmetry.

\subsection{Effective potential}\label{s:effectivepotential}
We briefly introduce the one-loop-corrected effective potential $V$, which is given by
\begin{align}
V = V_{\text{tree}} + V_\mathrm{CW} + V_\mathrm{CT} + V_T,
\label{eq:totalpotential}
\end{align}
where $V_{\text{tree}}$ is the tree-level potential of Eq.~\eqref{eq:v2hdm}, $V_\mathrm{CW}$ denotes the temperature-independent one-loop Coleman-Weinberg potential, $V_\mathrm{CT}$ indicates the counterterm potential, and $V_T$ accounts for the thermal corrections at finite temperature, respectively. Explicit expressions for $V_\mathrm{CW}$, $V_\mathrm{CT}$, and $V_T$ are given e.g.\ in \cite{Basler:2016obg,Basler:2018cwe}. In the following, we want to focus on the temperature-dependent part of the effective potential in the limit of high temperatures to discuss the possibility of \mbox{(non-)restoration} of the EW symmetry for temperatures $T\to \infty$.

The thermal contributions to the potential can be written as~\cite{Dolan:1973qd}
\begin{equation}\label{eq:thermalpotential}
	V_T = \sum_k n_k \frac{T^4}{2\pi^2} J_{\pm}^{(k)}\left(\frac{m_k^2}{T^2}\right)\,,
\end{equation}
where the sum extends over the fields $k=H^\pm, h, H, A, W_T, Z_T, W_L, Z_L, \gamma_L, t, b, \tau$ and $n_k$ denotes the number of degrees of freedom which is $n_k=2, 1, 1, 1, 4,2,2, 1, 1, 12, 12, 4$, respectively. 
The $J_+\left(\frac{m_k^2}{T^2}\right)$ and $J_-\left(\frac{m_k^2}{T^2}\right)$ are the thermal integrals for fermions and bosons, respectively,
\begin{equation}\label{eq:thermalintegral}
	J_\pm\left(\frac{m_k^2}{T^2}\right) = \mp\int_0^\infty\mathrm{d}y\,y^2\log\left[1 \pm e^{-\sqrt{y^2 + m_k^2/T^2}}\right].
\end{equation}
At high temperatures, the perturbative expansion used to obtain the above one-loop effective potential becomes unreliable due to higher-order terms not being suppressed~\cite{Quiros:1999jp}. To again restore the predictive power of the perturbation series, we include contributions from the resummation of diagrams of higher loop order, the so-called daisy diagrams. In this paper, we adopt the `Arnold-Espinosa' resummation method \cite{Arnold:1992rz} for taking into account the daisy diagrams\footnote{A different method to implement the thermal masses is given by the 'Parwani' method \cite{Parwani:1991gq}. The two approaches differ in the organization of the perturbative series. While the ’Arnold-Espinosa’ method consistently implements the thermal masses at one-loop level in the high-temperature expansion, the ’Parwani’ method admixes higher-order contributions, which at one-loop level could lead to dangerous artefacts.}. In practice, these contributions can be included by making the replacements
\begin{equation}\label{eq:thermalfunction}
	J_\pm^{(k)} =
		\begin{cases}
			J_-\left(\frac{m_k^2}{T^2}\right) - \dfrac{\pi}{6}\left[\left(\frac{\overline m_k^2}{T^2}\right)^{\frac32} - \left(\frac{m_k^2}{T^2}\right)^{\frac32}\right] & k = H^\pm, h, H, A, W_L, Z_L, \gamma_L\\[1em]
			J_-\left(\frac{m_k^2}{T^2}\right) & k = W_T, Z_T\\[1em]
			J_+\left(\frac{m_k^2}{T^2}\right) & k = t, b, \tau
		\end{cases}.
\end{equation}
Note that only the scalar Higgs fields and the longitudinal degrees of freedom of the gauge bosons receive daisy corrections, while the transversal degrees of freedom of the gauge bosons as well as the fermions do not. The $m_k$ in Eqs.~\eqref{eq:thermalpotential}--\eqref{eq:thermalintegral} denote the tree-level masses of the fields that depend implicitly on the temperature through the VEVs $\bar{\omega}_i = \bar{\omega}_i(T)$ as discussed in Sec.~\ref{s:cpc2hdm}, which are determined by minimising the loop-corrected effective potential at the temperature $T$. The $\overline{m}_k$ include thermal Debye corrections and therefore depend explicitly on $T$. Analytic expressions for the masses including thermal corrections can be found in Appendix~\ref{s:app:2hdmmasses}.

In the limit of high temperatures, the thermal integrals of Eq.~\eqref{eq:thermalintegral} can be approximated by \cite{Cline:1996mga}
\begin{align}
	J_+(x^2) &\stackrel{T\to\infty}{=} -\frac{7\pi^4}{360} + \frac{\pi^2}{24} x^2 + \mathcal{O}(x^4)\,,\\
	J_-(x^2) &\stackrel{T\to\infty}{=} -\frac{\pi^4}{45} + \frac{\pi^2}{12} x^2 - \frac{\pi}{6}(x^2)^{\frac32} + \mathcal{O}(x^4)\,,
\end{align}
with $x \equiv m_k/T$, where we only keep the leading terms in $T$ relevant for the following discussion. To summarise, we can approximate the thermal corrections to the effective potential in the high-temperature limit as\footnote{In the expansion of the thermal integrals $J_{\pm}(x^2)$ for small and large $x$ (i.e.\ high and low temperatures, respectively) as in \cite{Cline:1996mga}, a finite shift has to be added to smoothly connect the two regions. In the following discussion about EW symmetry restoration in the next section, this finite term will, however, not be relevant and we thus neglect it here.}
\begin{equation}\label{eq:thermalvapprox}
	V_T \stackrel{T\to\infty}{\approx} -\sum_k n_k
		\begin{cases}
			\displaystyle\frac{\pi^2}{90} T^4 - \frac{1}{24} m_k^2 T^2 + \frac{1}{12\pi}\overline m_k^3 T & k = H^\pm, h, H, A, W_L, Z_L, \gamma_L\\[1em]
			\displaystyle\frac{\pi^2}{90} T^4 - \frac{1}{24} m_k^2 T^2 + \frac{1}{12\pi} m_k^3 T & k = W_T, Z_T\\[1em]
			\displaystyle\frac{7\pi^2}{720} T^4 - \frac{1}{48} m_k^2 T^2 & k = t, b, \tau
		\end{cases}.
\end{equation}

\subsection{Symmetry non-restoration}\label{s:NonRestoration}
Recently, several models have been studied where at high temperatures, the electroweak symmetry is not restored, see e.g.~Refs.~\cite{Ramsey-Musolf:2017tgh,Meade:2018saz,Baldes:2018nel,Glioti:2018roy,Matsedonskyi:2020mlz,Carena:2021onl,Biekotter:2021ysx,Biekotter:2022kgf}. In this case, the origin of the potential is not the global minimum at high temperature. The phenomenon can occur even in the SM when the scalar quartic coupling is large, as we verified explicitly.

In order to analyse the situation in the 2HDM, we study the curvature of the effective potential with respect to the electroweak VEVs around the origin, i.e.\ we calculate the Hessian matrix at the origin,
\begin{equation}\label{eq:hessianveff}
	H_{ij} \equiv \frac{\partial^2 V_T}{\partial\bar{\omega}_i \partial\bar{\omega}_j}\bigg|_{\bar{\omega}_{i,j}=0}, \quad i,j=1,2.
\end{equation}
Note that in this discussion, we limit ourselves to the thermal part of the one-loop effective potential, as we are only interested in the behaviour for $T\to\infty$ and the temperature-independent parts thus do not play a role. The leading term of $H_{ij}$ in $T$, which is $\propto T^2$, determines the behaviour for $T\to\infty$. We therefore evaluate $H_{ij}/T^2$ for $T\to\infty$. We find that this expression takes the following form,
\begin{equation}
	\mathcal{H} \equiv \lim_{T\to\infty} \frac{H}{T^2} = \lim_{T\to\infty}
		\begin{pmatrix}
			\frac{H_{11}}{T^2} & \frac{H_{12}}{T^2} \\
			\frac{H_{21}}{T^2} & \frac{H_{22}}{T^2}
		\end{pmatrix}
        =
		\begin{pmatrix}
			\mathcal{H}_{11} & 0 \\
			0 & \mathcal{H}_{22}
		\end{pmatrix},
\end{equation}
with
\begin{equation}\label{eq:symrestoreAEcoeffs}
	\begin{split}
		\mathcal{H}_{11} &= c_1 - \frac{1}{16\pi}\left[\sqrt{2}\left(3g^3 + g'^3\right) + 4\left(3\sqrt{c_1} \lambda_1 + \sqrt{c_2}\left(2\lambda_3 + \lambda_4\right)\right)\right]\,,\\
		\mathcal{H}_{22} &= c_2 - \frac{1}{16\pi}\left[\sqrt{2}\left(3g^3 + g'^3\right) + 4\left(3\sqrt{c_2} \lambda_2 + \sqrt{c_1}\left(2\lambda_3 + \lambda_4\right)\right)\right]\,,
	\end{split}
\end{equation}
and with $c_1$ and $c_2$ given by Eqs.~\eqref{eq:c1coeff} and \eqref{eq:c2coeff}, respectively. In order for the stationary point at the origin to be a (local or global) minimum, all eigenvalues of the Hessian matrix are required to be positive. We therefore obtain the following conditions for a minimum at the origin,
\begin{equation}\label{eq:symnonresconditions}
	\mathcal{H}_{11} > 0 \qquad\text{and}\qquad \mathcal{H}_{22} > 0\,.
\end{equation}
If the stationary point is a global minimum, the EW symmetry is restored for $T\to\infty$; if it is, however, only a local minimum (with a finite probability to tunnel into the deeper global minimum), or if any of the two inequalities of Eq.~\eqref{eq:symnonresconditions} is not fulfilled, we observe non-restoration of the EW symmetry at $T\to\infty$.

\begin{figure}[tp]
	\centering
	\includegraphics[width=.5\textwidth]{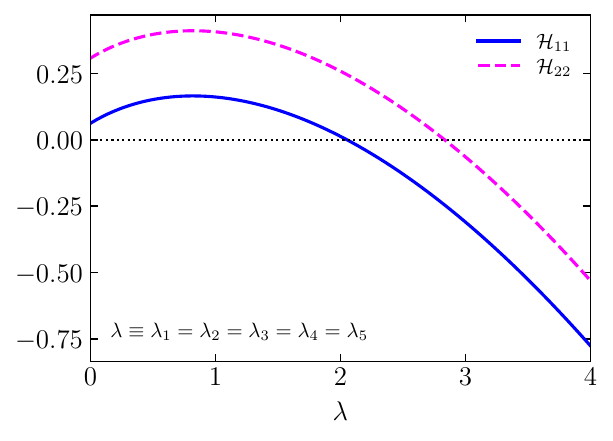}\hfill\includegraphics[width=.5\textwidth]{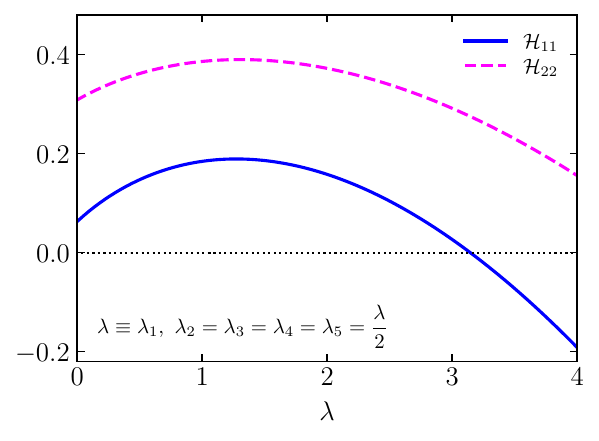}
	\caption{Curvatures ${\cal H}_{11}$ (lower/solid blue) and ${\cal H}_{22}$ (upper/dashed magenta) at high-temperatures, as defined in Eq.~\eqref{eq:symrestoreAEcoeffs}. Left: all scalar couplings are set equal, $\lambda \equiv \lambda_1 = \lambda_2 = \lambda_3 = \lambda_4 = \lambda_5$. Right: $\lambda_1$ is twice as large as the other couplings (which are again set equal), $\lambda \equiv \lambda_1$ and $\lambda_2 = \lambda_3 = \lambda_4 = \lambda_5 = \frac{\lambda}{2}$. The remaining coupling values are given in Eq.~(\ref{eq:coupvalues}).} 
	\label{fig:lambdamax}
\end{figure}
In Fig.~\ref{fig:lambdamax}, we plot ${\cal H}_{11}$ and ${\cal H}_{22}$ for two simple cases, one where all $\lambda \equiv \lambda_1 = \lambda_2 = \lambda_3 = \lambda_4 = \lambda_5$ values are set equal (left plot), and one where $\lambda_1$ is larger than all the other $\lambda_{2,3,4,5}$, more specifically $\lambda \equiv \lambda_1$ and $\lambda_2 = \lambda_3 = \lambda_4 = \lambda_5 = \frac{\lambda}{2}$. All other couplings are given by 
\begin{equation}
    g = 0.653,\quad g' = 0.350,\quad y_t = 0.991,\quad y_b = 0.03,\quad y_\tau = 0.01\,.
\label{eq:coupvalues}
\end{equation}
Electroweak symmetry restoration is only possible if both lines are above zero. The limit on the scalar couplings up to which a restoration is still possible can therefore be read off as the root of the lower of the two lines (which in both of our cases is the blue line corresponding to $\mathcal{H}_{11}$). It can be seen that in the first case, there is the possibility for restoring the EW symmetry at high temperatures only for $\lambda < 2$, while in the second case, higher values are allowed with $\lambda \lesssim 3$. Thus there is an upper limit on the quartic couplings up to which EW symmetry can be restored. The exact values depend on the chosen configuration of the parameters.

%% file: 5-numerics.tex
\section{Numerical analysis}\label{s:numerics}
We now turn to the numerical analysis. We want to investigate to which extent the 2HDM can exhibit a charge-breaking phase at non-zero temperature when we apply the full one-loop treatment of the effective potential and we require in addition that all relevant theoretical and experimental constraints are fulfilled. We will also analyse how this relates to EW symmetry non-restoration. Note that the $m_{12}^2$ parameter is chosen non-zero in our numerical analysis. We remind the reader that we are considering the 2HDM type I.

\subsection{Set-up of the scan and constraints}\label{s:numerics:setup}
\begin{table}[tp]
    \centering
    \renewcommand{\arraystretch}{1.2}
    \begin{tabular}{cc}
        \toprule
        Parameter & Scan range\\\midrule
        $\lambda_1, \lambda_2, \lambda_3, \lambda_4$ & $[0, 4\pi]$\\
        $\lambda_5$ & $[-4\pi, 4\pi]$\\
        $m_{11}^2, m_{22}^2$ & $[-10^6, 0]$ GeV$^2$\\
        $m_{12}^2$ & $[0, 10^6]$ GeV$^2$\\
        \bottomrule
    \end{tabular}
    \caption{Scan ranges of the parameters of the 2HDM potential of Eq.~(\ref{eq:v2hdm}).}
    \label{tab:paramscanranges}
\end{table}
In order to find parameter points that are viable in the sense that they satisfy the applied constraints, which we describe here below, we perform a scan in the input parameters of the model. For this we use in-house \texttt{Python} routines with an implementation of the CP-conserving 2HDM and its tree-level potential. The scan ranges of the parameters are given in Table~\ref{tab:paramscanranges}.

Since an initial random scan in the parameter space did not produce any points which exhibit a CB phase at non-zero temperature for the full one-loop effective potential and which are additionally compatible with all required constraints, we had to improve our scanning method. This is done by starting from \emph{seed points} based on the simple high-$T$ approximation outlined in Sec.~\ref{s:chargebreaking}. They are defined as follows:\footnote{Let us remark here, that this scan method does not bias our scan, but rather is an informed procedure to efficiently search for parameter points exhibiting an intermediate CB phase. We explicitly checked this through an extensive random scan which did not find valid points in regions of the parameter space different from the one found in the seed-point induced scan.}

We require the temperature-independent conditions for a CB minimum of Eq.~\eqref{CB-condition-1} to be fulfilled (as we are interested in the possibility of a CB phase at non-zero temperature). The temperature-dependent condition Eqs.~(\ref{CB-condition-3}) and (\ref{CB-condition-4}), however, must not be fulfilled at $T = 0$. Assuming $\lambda_3 > 0$, we impose the conditions Eq.~\eqref{toy-model-condition} on the quadratic mass parameters to intelligently generate points that lie in suitable regions of the phase diagram, i.e.\ outside of the blue CB region (cf. Fig.~\ref{fig-regions-1}) and which will always have a trajectory exhibiting an intermediate CB phase
(cf.~Fig.~\ref{fig-regions-2}) for the toy model of Sec.~\ref{s:chargebreaking}. The resulting points are then considered in the following as \emph{seed points} for the appearance of a CB minimum at $T > 0$. Note that due to $m_{12}^2 > 0$ in the model of Sec.~\ref{s:cpc2hdm}, it is possible that using the tree-level potential and the high-$T$ approximation for the evolution of the quadratic parameters only, the trajectory misses the CB phase, as discussed in Sec.~\ref{s:chargebreaking:softz2hdm}. Nevertheless, we do not discard these points as in the full one-loop treatment of the temperature dependence of the effective potential including the temperature-corrected mass terms, it is possible that the CB phase is recovered. Since these seed points are based on the derivation for the tree-level potential with the high-$T$ approximation, it can also be that not all seed points lead to charge-breaking minima when considering the complete temperature-dependent potential. They are a good starting point, however, to find suitable scenarios. 

Before applying any constraints, we rescaled our parameters such that we obtained the correct numerical values for the SM-like CP-even Higgs mass, $m_h = 125.09$ GeV, and the vacuum expectation value, $v = 246.22$ GeV, i.e.
\begin{equation}
    m_{ij}^2 ~\rightarrow~ m_{ij}^2\frac{m_h^2}{m_{h,0}^2},~~ i,j =1,2, \qquad
\lambda_k ~\rightarrow~ \lambda_k\frac{m_h^2}{m_{h,0}^2}\frac{v_0^2}{v^2},~~ k = 1,..., 5,
\end{equation}
where $v_0$ and $m_{h,0}$ are the VEV and the lightest neutral CP-even scalar mass, respectively, as obtained from the parameters before the rescaling. 
In our scan, we define the mixing angle $\alpha$ that diagonalises the Higgs mass matrix for the CP-even states such that the lightest CP-even mass eigenstate always corresponds to the SM-like Higgs boson with mass $m_h$.

We apply the following theoretical constraints on our parameter samples:
\begin{itemize}
	\item The potential is required to be bounded from below (see Eq.~\eqref{BFB}).
    \item The quartic couplings have to fulfill perturbativity. More specifically, we demand $|\lambda_i| < 4\pi$ ($i=1,...,5$).
	\item We demand that tree-level unitarity is preserved \cite{Akeroyd:2000wc,Ginzburg:2003fe}.
	\item The neutral CP-even tree-level minimum is required to be the global one, which can be tested through a simple condition \cite{Barroso:2013awa}.
\item We furthermore dismiss all points that, at tree level, do not exhibit a neutral vacuum at $T = 0$. 
\end{itemize}

On the resulting points, we finally employ the following experimental constraints, using the program \ScannerS{}~\cite{Coimbra:2013qq,Ferreira:2014dya,Costa:2015llh,Muhlleitner:2016mzt,Muhlleitner:2020wwk}: 
\begin{itemize}
	\item We demand the 125~GeV Higgs boson to behave SM-like. Compatibility with the Higgs signal data is checked through \texttt{HiggsSignals} \cite{Bechtle:2013xfa}, which is linked to \texttt{ScannerS}.
 We require 95\% C.L.\ exclusion limits on non-observed scalar states by using \texttt{HiggsBounds} \cite{Bechtle:2008jh,Bechtle:2011sb,Bechtle:2013wla}.
 \item We impose compatibility with the electroweak precision data by demanding the computed $S$, $T$ and $U$ parameter values \cite{Peskin:1991sw} to be within 2$\sigma$ of the SM fit \cite{STU3}, taking into account the full correlation among the three parameters.
	\item Consistency with recent flavour constraints is ensured by testing for the compatibility with ${\cal R}_b$ \cite{Rb1,Rb2} and $B \to X_s \gamma$ \cite{flavor1,
Hermann:2012fc,Misiak:2015xwa,Misiak:2017bgg,Misiak:2020vlo} in the $m_{H^\pm}$-$\tan \beta$ plane.
\end{itemize}

For the thus obtained seed points (obtained with the approximate high-temperature method outlined above) we then checked if they exhibit intermediate CB phases using the full one-loop effective potential. This is done with the program \BSMPT{}~\texttt{v2.6.0}~\cite{Basler:2018cwe,Basler:2020nrq} which performs the minimisation of the one-loop effective potential given in Sec.~\ref{s:effectivepotential}. Here, we make use of the \texttt{R2HDM} model implementation.

\subsection{Parameter dependence}\label{s:numerics:paramsearch}
We now present the results of our parameter scan, and discuss the interplay between the constraints, the appearance of a CB phase, and the restoration of the EW symmetry. Our procedure here is as follows:

As we found that it was the hardest to fulfill simultaneously the requirements for a CB phase and the Higgs signal constraints, we first generated about 5000 seed points as outlined above while only enforcing the experimental $S$, $T$, and $U$ parameter as well as the flavour constraints. For these seed points, the signal strength of the SM-like Higgs coupling to two photons, $\mu_{\gamma\gamma}$, was typically significantly below one. Thus, in order to additionally fulfill the Higgs signal constraints, we varied the parameters of these starting points in the direction of increasing $\mu_{\gamma\gamma}$ until the Higgs signal constraints as checked by \texttt{ScannerS} were satisfied (together with all other applied experimental constraints). During the course of our analysis, we found that points exhibiting an intermediate CB phase in the high-$T$ approximation cannot be made compatible with the experimental constraints as they lead to $\mu_{\gamma\gamma}$ values below the experimental limit. The reason is that a CB phase obtained in this approximation implies too light charged Higgs masses. Note, that all points generated in the scan lead to the correct EW minimum at zero temperature using the full one-loop corrections in the effective potential.

\begin{figure}[t]
	\centering
	\begin{tabular}{cc}
		\includegraphics[width=.47\textwidth]{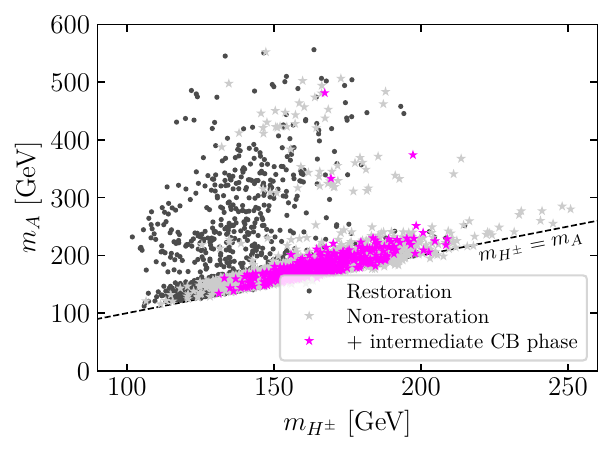} & \includegraphics[width=.47\textwidth]{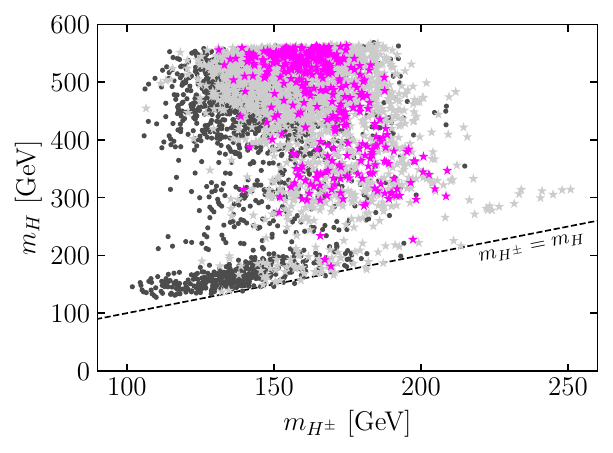} \\
		(a) & (b) \\
		\includegraphics[width=.47\textwidth]{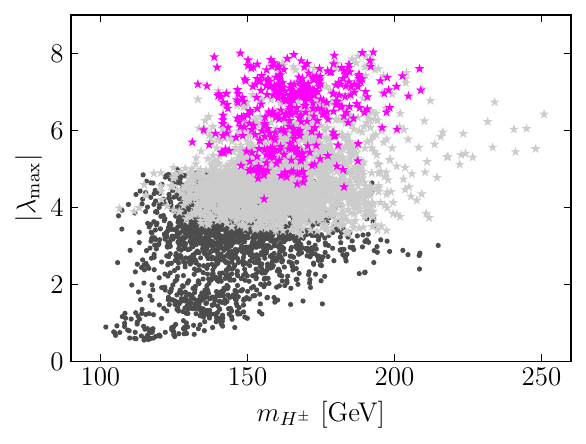} & \includegraphics[width=.5\textwidth]{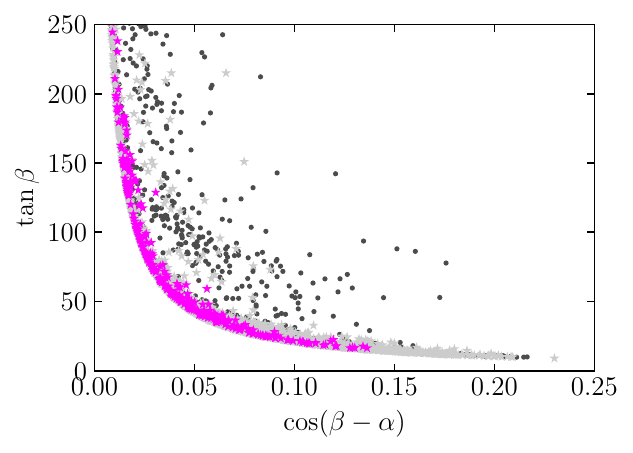} \\
		(c) & (d)
	\end{tabular}
	\caption{Parameter points of our scan shown for different parameter planes as discussed in the text. All points fulfill the experimental constraints from \ScannerS{}. Dark (light) grey points: EW symmetry (non-)restoration at high temperatures. Magenta points exhibit in addition an intermediate CB phase at non-zero temperature using the one-loop effective potential.}
	\label{fig:scan2figures}
\end{figure}

In Fig.~\ref{fig:scan2figures}, we show the results of our scan for the following set of plots displaying different parameter combinations,
\begin{enumerate}[label=(\alph*)]
	\item top left: charged Higgs mass $m_{H^\pm}$ vs.\ CP-odd Higgs mass $m_A$;
	\item top right: charged Higgs mass $m_{H^\pm}$ vs.\ heavy CP-even Higgs mass $m_H$;
	\item bottom left: charged Higgs mass $m_{H^\pm}$ vs.\ $|\lambda_{\text{max}}| \equiv\max(|\lambda_1|, |\lambda_2|, |\lambda_3|, |\lambda_4|, |\lambda_5|)$;
	\item bottom right: Higgs-gauge coupling modifier $\cos(\beta - \alpha)$ vs.\ $\tan\beta$.
\end{enumerate}
In all of the plots, points for which the restoration of the EW symmetry at high temperatures is possible\footnote{For clarity, we again emphasise that even if the conditions of Eq.~\eqref{eq:symnonresconditions} are fulfilled, the EW symmetry could still be broken in the high-temperature limit if the minimum at the origin is not the global one.} are coloured in dark-grey, while those with non-restoration of the EW symmetry are shown in light-grey. Those points which additionally exhibit an intermediate charge-breaking phase at non-zero temperature are coloured in magenta.

We first discuss the impact of the experimental constraints and symmetry (non-)restoration on the parameter regions returned by the scan. As can be inferred from Fig.~\ref{fig:scan2figures}~(a), the totality of the light and dark grey points features charged Higgs mass values above 100~GeV. This is due to the application of the experimental constraints. From Figs.~\ref{fig:scan2figures}~(a) and (b) in the top row, it can be seen that the majority of the points favour a degenerate configuration of the charged and CP-odd or heavy CP-even scalar masses, as required by the $S$, $T$, $U$-precision observables.\footnote{For those points in Figs.~\ref{fig:scan2figures}~(a) and (b), when $m_A$ $(m_H)$ is not degenerate with $m_{H^\pm}$, it is the $m_H$ $(m_A)$ mass that becomes degenerate then.} Fig.~\ref{fig:scan2figures}~(c) shows that only points with $|\lambda_{\text{max}}| \lsim 5$ lead to EW symmetry restoration. There are no further dark grey points behind the light grey ones for $|\lambda_{\text{max}}| \gsim 5$. The appearance of an upper limit is in accordance with our discussion in Sec.~\ref{s:NonRestoration}. In Fig.~\ref{fig:scan2figures}~(d), we observe that due to the experimental constraints, only points with $\cos(\beta-\alpha) \lsim 0.23$ are allowed. We applied here among others the limits from \cite{ATLAS:2020note}. Note, however, that the recent analysis \cite{ATLAS:2021vrm} constrains $\cos(\beta-\alpha)$ to be close to 0 for the 2HDM type I, with only small non-zero values allowed for larger values of $\tan\beta$.\footnote{The asymmetric shape of the allowed region in $\cos(\beta-\alpha)$ is driven by recent analyses like the $b\bar{b}$ final state in associated production of the Higgs boson with a top quark pair \cite{ATLAS:2021qou}, which is dominated by systematic uncertainties \cite{Biekotter:2022ckj}.}

We now additionally require an intermediate CB phase for the one-loop effective potential, which is fulfilled by the magenta points in Fig.~\ref{fig:scan2figures}. The number of viable points is significantly reduced, down to approximately 500 points. None of these remaining points exhibit EW symmetry restoration, which is also reflected by the fact that we have a relatively large maximum scalar coupling, $4 \lsim |\lambda_{\text{max}}| \lsim 8$, which we typically find to be $\lambda_1$. When we relax the experimental constraints, however, points are found that exhibit a CB phase and simultaneously allow for EW symmetry restoration. We checked this explicitly in a preliminary step of our scan. The appearance of a CB phase is hence in tension with two constraints that cannot be reconciled simultaneously: It favours rather large scalar quartic couplings that hamper electroweak symmetry restoration. On the other hand, it requires relatively low charged Higgs masses, which puts tension on compatibility with the LHC constraints.

The phenomenological implications for the collider physics are as follows: The bulk of the points shows a mass degeneracy $m_{H^\pm} \approx m_A$, which corresponds to $\lambda_4 \approx \lambda_5$ (see Eqs.~\eqref{eq:chargedhiggsmass} and \eqref{eq:cpoddhiggsmass}). The neutral heavy Higgs mass on the other hand is, for most of the parameter points, separated from $m_A$ and $m_{H^\pm}$ by a gap of $\gsim 100$~GeV, which would allow for scalar decays into mixed scalar and gauge boson final states, $H \to AZ$ or $H \to H^\pm W^\mp$. Future increased sensitivity in the $H\to AZ$ search channels (cf.~e.g.~\cite{CMS:2019ogx,ATLAS:2020gxx} for existing searches in the $ZA$ final state) and future searches in the $W^\pm H^\mp$ final state, where no experimental searches exist yet, could hence either confirm these parameter points or further constrain them. Applying the loop-corrected effective potential to find a CB phase, which is simultaneously compatible with experimental constraints, also reduces the maximum possible value of the charged Higgs mass to below $\sim$210~GeV and it drives the model further towards the SM limit, as now $\cos(\beta-\alpha)\lsim 0.14$.

For completeness, we present in App.~\ref{s:app:lambdas} also the parameter distributions in the quartic couplings $\lambda_1$ to $\lambda_5$ and in the squared  soft-breaking mass parameter $m_{12}^2$ as a function of the charged Higgs mass, respectively, as well as $m_{22}^2$ versus $m_{11}^2$.  As can be inferred from the figures, for most of the points, either $\lambda_1$ or $\lambda_4$ is the largest self coupling, while for the points exhibiting an intermediate CB phase, it is almost exclusively $\lambda_1$. The soft breaking parameter $m_{12}^2$ is reduced to somewhat smaller values due to the relatively small charged Higgs mass when an intermediate CB phase is required. The values of $m_{11}^2$ and $m_{22}^2$ are negative as required for successful generation of a CB phase, and the $m_{22}^2$ value is rather constrained through the measured Higgs mass value (cf.~Sec.~\ref{s:chargebreaking}). We remark that the parameters that we presented in our figures denote the full set of input parameters for the real 2HDM, including also the parameters $v$ and $m_h$ which are fixed to the SM values, so that all possible desired observables can in principle be derived from the delivered information.

We finally note that we have additionally checked in a separate scan that when requiring the parameter points to fulfill all three classes of constraints simultaneously, i.e.\ the experimental constraints, an intermediate CB phase for the one-loop effective potential, and the possibility for EW symmetry restoration at high temperatures, no viable points are found.

\subsection{Temperature evolution}\label{s:numerics:tevo}
\begin{table}[tp]
    \centering
    \renewcommand{\arraystretch}{1.2}
    \begin{tabular}{ccccccc}
        \toprule
        & $m_H$ [GeV] & $m_A$ [GeV] & $m_{H^\pm}$ [GeV] & $\tan\beta$ & $\cos(\beta - \alpha)$ & $m_{12}^2$ [GeV$^2$] \\\midrule
        \texttt{BP1} & 562.84 & 168.56 & 164.51 & 16.58 & 0.128 & 18933.44 \\
        \texttt{BP2} & 548.51 & 175.55 & 171.43 & 45.32 & 0.048 & 6629.97 \\
        \texttt{BP3} & 342.52 & 230.02 & 183.72 & 286.00 & 0.009 & 410.17 \\
        \texttt{BP4} & 558.56 & 194.52 & 168.43 & 80.84 & 0.026 & 3857.90 \\
        \bottomrule
    \end{tabular}
    \caption{Selected benchmark points from the parameter scan with the constraints as in Fig.~\ref{fig:scan2figures}.}
    \label{tab:benchmarkpoints}
\end{table}
In the following, we discuss the temperature evolution of the absolute values of the EW and CB VEVs $\bar\omega_1$, $\bar\omega_2$, $\bar\omega_\CB$ of the scalar fields in Eqs.~(\ref{eq:phi1}) and (\ref{eq:phi2}) for several benchmark points that have been selected due to different features that they exhibit in the evolution of the VEVs. They have been selected from the set of magenta points of Fig.~\ref{fig:scan2figures}, i.e.\ without demanding EW symmetry restoration at high temperatures, but requiring a CB phase using the one-loop effective potential\footnote{In the whole sample of points of Fig.~\ref{fig:scan2figures}, the CP breaking phase is found to be zero across the scanned temperature range, i.e.\ we do not observe spontaneous CP violation.}.
The parameter values of the chosen benchmark points, given in terms of the physical masses and mixing angles, are listed in Table~\ref{tab:benchmarkpoints}.

\begin{figure}[tp]
	\centering
    \begin{tabular}{cc}
	    \includegraphics[width=.48\textwidth]{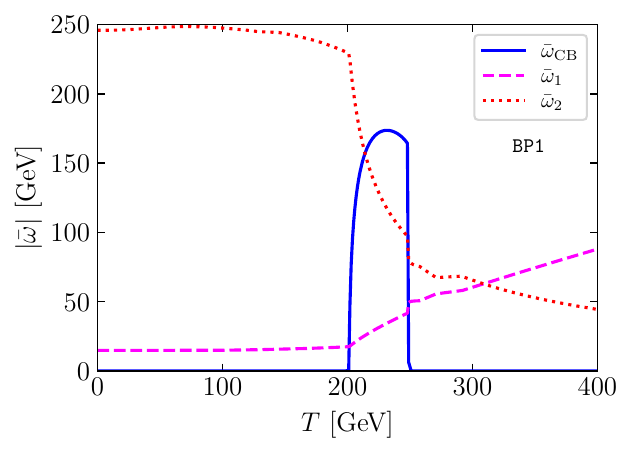} & \includegraphics[width=.48\textwidth]{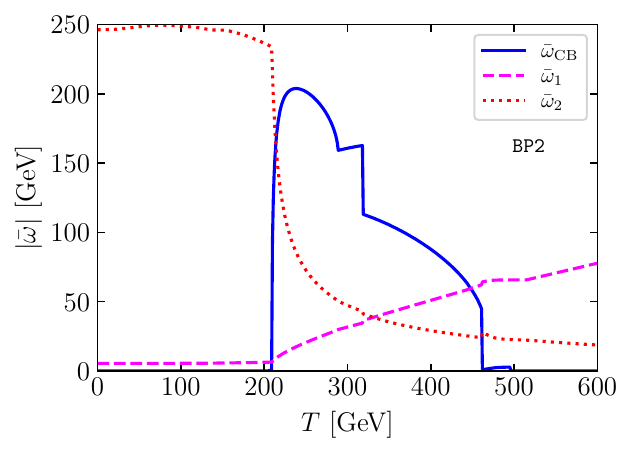} \\
        (a) & (b) \\
        \includegraphics[width=.48\textwidth]{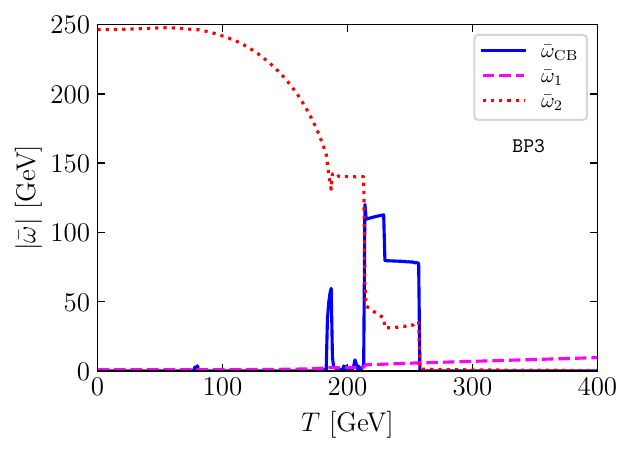} & \includegraphics[width=.48\textwidth]{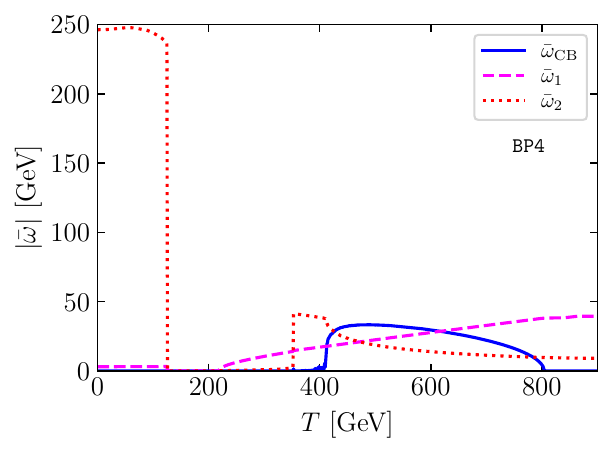} \\
        (c) & (d)
    \end{tabular}
	\caption{Absolute values of the EW and CB VEVs $\bar{\omega}_{1}$ (magenta dashed) $\bar{\omega}_2$ (orange dotted) and $\bar{\omega}_{\text{CB}}$ (blue solid) as a function of the temperature $T$, calculated using the \BSMPT{} code, for the selected benchmark points of Table~\ref{tab:benchmarkpoints}: \texttt{BP1} (a, top left), \texttt{BP2} (b, top right), \texttt{BP3} (c, bottom left), and \texttt{BP4} (d, bottom right).}
	\label{fig:scan2vevevo}
\end{figure}
The temperature evolution is shown in Fig.~\ref{fig:scan2vevevo} which we discuss in more detail now. First of all, the plots show that we find CB phases (cf.~full blue lines) of different durations and magnitudes. The benchmark point \texttt{BP1}, Fig.~\ref{fig:scan2vevevo}~(a), displays the following sequence of phases (from high temperatures to the EW vacuum with a VEV of 246~GeV at $T = 0$):
\begin{center}
    Neutral $\to$ CB $\to$ EW vacuum.
\end{center}
The intermediate CB phase starts around $T = 250$ GeV with a duration of approximately $\Delta T = 50$ GeV and a pronounced maximum of $|\bar\omega_\CB| \approx 170$~GeV. We find a first-order phase transition from the electrically neutral to the CB phase, while the second phase transition from the CB phase to $\bar{\omega}_{\text{CB}}=0$ at $T = 200$~GeV is of second order\footnote{We call a phase transition ``first order'' if the corresponding VEV shows a jump in its value at the considered temperature. We call it ``strong first order'' if the jump of the VEV divided by the temperature is larger than one. We call the phase transition ``second order'' if it continuously changes its value across a non-zero temperature interval.}. In parallel, over the shown temperature range, the EW VEV $\bar{\omega}_1$
decreases, while $\bar{\omega}_2$ increases with falling temperature. They both show a first-order phase transition at the start of the non-zero CB phase at $T = 250$~GeV. Afterwards $\bar{\omega}_2$ increases (strongly) and $\bar{\omega}_1$ continues to decrease (slowly), a behaviour that is continued by each of the VEVs at the second CB phase transition at $T = 200$~GeV but with a different slope. They reach $v = \sqrt{v_1^2+v_2^2}= 246$~GeV at $T=0$, as expected and required. The symmetry non-restoration of the EW VEV at high temperatures is clearly visible.

The phase evolution for \texttt{BP2} displayed in Fig.~\ref{fig:scan2vevevo}~(b) shows the same qualitative behaviour as \texttt{BP1}, but with a longer duration of the intermediate CB phase, which amounts to approximately $\Delta T = 300$ GeV. Additionally, we observe several phase transitions within the CB phase, which starts around 480~GeV. They lie at $T = 460$ GeV, $310$ GeV, and 280~GeV, respectively, and according to our definition are of first order, first order, and of second order, respectively. 

Benchmark point \texttt{BP3} features the largest mass splitting $m_A - m_{H^\pm} \approx 50$ GeV of all selected benchmark points and a very large value of $\tan\beta$. As can be inferred from Fig.~\ref{fig:scan2vevevo}~(c) it provides the possibility of having more than one distinct CB phase. We find as the sequence of phases\footnote{The wiggles and spikes are numerical instabilities in the determination of the global minimum, when e.g.\ different minima are almost degenerate.}:
\begin{center}
    Neutral $\to$ CB $\to$ neutral $\to$ CB $\to$ EW vacuum.
\end{center}
While the first CB phase, starting at around $T = 260$ GeV, has a duration of approximately $\Delta T = 50$ GeV, the second CB phase is much shorter with approximately $\Delta T = 5$ GeV. We again find a combination of first- and second-order phase transitions, and an additional phase transition within the CB phase at around $T = 230$ GeV. The EW VEVs show the same qualitative behaviour as in the previous benchmark points with several intermediate phase transitions at the same temperatures as the CB ones and of first or second order.

While for \texttt{BP1}--\texttt{BP3}, the EW symmetry is not restored at all, for several points, we find an intermediate restoration of the EW symmetry. We selected \texttt{BP4} among these points and show the phase evolution in Fig.~\ref{fig:scan2vevevo}~(d). The sequence of transitions is:
\begin{center}
    Neutral $\to$ CB $\to$ neutral $\to$ neutral $\to$ EW symmetric $\to$ EW vacuum.
\end{center}
The EW symmetric phase starts at around $T = 220$ GeV and persists down to approximately $T = 120$ GeV. We have a second neutral phase in between the CB phase and the EW symmetric phase, as in the second neutral phase right before the EW symmetric phase, the VEV $\bar\omega_2$ becomes very small and almost vanishes. We find a long CB phase starting at $T = 800$ GeV with approximately $\Delta T = 400$ GeV. The phase transition from the EW symmetric to the broken phase with the EW vacuum is of first order.

To summarise, we confirm the existence of intermediate CB phases in the CP-conserving 2HDM not only in the high-temperature approximation as previously discussed in the literature \cite{Ivanov:2008er,Ginzburg:2009dp}, but also using the one-loop-corrected effective potential with thermal corrections. While the existence of a CB phase in the high-temperature approximation generally requires a small charged Higgs mass $m_{H^\pm} \lsim 100$ GeV which is in tension with experimental searches, we find viable parameter points with larger charged scalar masses which are in agreement with current experimental limits by requiring a CB phase using the full one-loop effective potential with thermal corrections, not relying on the high-temperature approximation only. However, as discussed in Sec.~\ref{s:chargebreaking} in the context of the toy model and the high-temperature approximation, reasonably large scalar couplings are required for a CB phase to occur, which is also found to be the case when using the one-loop effective potential for the  CP-conserving 2HDM. In our scan, the magnitudes of the scalar couplings are consequently in tension with the restoration of the EW symmetry at high temperatures according to the conditions derived in Sec.~\ref{s:symrestoration}, and we do not find any points that have a CB phase at non-zero temperature, fulfill all experimental constraints, and lead to EW symmetry restoration at high temperatures simultaneously. Nonetheless, we find for several of the parameter points, besides the appearance of an intermediate CB phase, interesting phase evolutions, including different sequences of phases with first- or second-order phase transitions and intermediate phases of EW symmetry restoration.

%% file: 6-conclusions.tex
\section{Conclusions and outlook}\label{s:conclusion}
The Higgs potential of extended Higgs sectors, which are motivated by open puzzles of the SM, show an interesting and complex vacuum structure. Travelling back in time to higher temperatures, the complexity of this structure can change and show rich and even exotic patterns of minima sequences that are acquired. In this paper we investigated the interesting question if the 2HDM type I could go through a CB phase on its way from the early universe to today's electroweak vacuum. Such an intermediate CB phase during the thermal evolution of the universe could be the seed of magnetic fields in the universe, whose origin is still unknown, and it could have interesting consequences for the cosmological evolution and the generation of dark matter. We took in this paper the first step and analysed if such a CB phase could take place at all and under which conditions.
While it was already shown in previous works, using a simpler high-temperature approximation for the potential, that a CB phase exists, we demonstrated that this still holds when taking into account the full one-loop corrected effective potential including thermal masses.

Our investigations showed, however, that a CB phase is in tension with two requirements, which can be reconciled individually but not simultaneously with a CB phase. The appearance of a CB phase favours rather large scalar quartic couplings, which works against electroweak symmetry restoration requiring small quartic couplings. A CB phase is found to require furthermore rather low charged Higgs masses, which puts tension on compatibility with the LHC constraints. The bottom line is that the CB phase can accommodate symmetry restoration at the price of non-compatibility with the experimental constraints. The appearance of a CB phase can also be in accordance with the experimental constraints but then at the price of symmetry non-restoration. If we accept symmetry non-restoration by assuming that at some higher temperature a new mechanism restores the EW symmetry, then we can derive typical collider features that are provoked by the requirement of a CB phase during the thermal evolution of the universe. These are found to be for our model, the 2HDM type I, maximal quartic couplings in the range between about 4 and 8, and rather light charged Higgs masses between about 130 and 210 GeV (this also explains why 2HDM type II models cannot accommodate a CB phase as here the charged Higgs mass is constrained to be above 800 GeV \cite{Hermann:2012fc,Misiak:2015xwa,Misiak:2017bgg,Misiak:2020vlo}). They are found to be mostly degenerate with the pseudoscalar mass, while the heavier neutral scalar mass is typically by more than 100 GeV heavier than the charged Higgs mass. Furthermore, the model is driven closer to the SM limit with $\cos(\beta-\alpha) \lsim 0.14$. This implies interesting LHC phenomenology as e.g.\ the possibility of $H \to W^\pm H^\mp$ decays.

Concerning the possible realisation of CB phases at non-zero temperature, performing a random scan within the a priori identified region of the parameter space favourable for such a situation, numerous scenarios can be found, albeit requiring a reasonably large scan sample. There is hence some degree of fine-tuning involved in finding the existence of intermediate CB phases in the 2HDM during the thermal evolution of the universe.

Accepting the absence of symmetry restoration at high temperatures, we chose four benchmark points with an intermediate CB phase which are compatible with the experimental constraints and investigated the temperature evolution of their VEVs. We found interesting phase histories, with CB phases of different durations and magnitudes and more or less complex sequences of consequent or parallel neutral and CB phases from first or second order phase transitions, including intermediate phases of EW symmetry restoration.

The analysis of the consequences of such thermal histories for cosmology, gravitational waves, dark matter, and also collider phenomenology is left for a separate dedicated study. Our analysis has shown, however, in this first important step that CB phases are possible and under which conditions they are possible. Our investigations furthermore once again demonstrate how powerful the investigation of the vacuum structure is in providing insights in beyond-the-SM physics and corners of the particle world where we do not have direct access.

\section*{Acknowledgements}
C.\ B.\ would like to thank Duarte Azevedo and Thomas Biekötter for valuable discussions.
The work of M.\ A.\ is supported in part by the Japan Society for the Promotion of Sciences Grant-in-Aid for Scientific Research (Grant No.~20H00160). 
The work of H.\ S.\ is supported by JST SPRING, Grand No.~JPMJSP2135. The work of C.\ B.\ and M.\ M.\ is supported by the DFG Collaborative Research Center TRR257 ``Particle Physics Phenomenology after the Higgs Discovery''. I.\ P.\ I.\ is supported by the Guangdong Basic and Applied Basic Research Fund through the project ``Evolution of the early Universe in the three-Higgs-doublet model''. Figures~\ref{fig-regions-1} and \ref{fig-regions-2} have been generated with the help of the Ti\emph{k}Z package~\cite{tikzmanual}. All plots in figures \ref{fig:lambdamax}--\ref{fig:scan2vevevo} have been created with the Python package \texttt{Matplotlib}~\cite{Hunter:2007}.

%% file: 7-appendix.tex
\section{Masses in the 2HDM}\label{s:app:2hdmmasses}
In this appendix, we show analytic expressions for the masses of all particles relevant for our discussion of the finite-temperature potential.

\subsection{Fermion masses}\label{s:app:2hdmmasses:fermions}
The fermion masses do not get Debye corrections and, assuming a 2HDM type-I, are therefore given by,
\begin{equation}
	m_f = \frac{y_f}{\sqrt{2}} \bar\omega_2\,,
\end{equation}
where the Yukawa coupling $y_f$ is defined via the VEV $v_2 = \bar\omega_2|_{T = 0}$ and the fermion mass $m_f(T = 0)$ at zero temperature,
\begin{equation}
    y_f = \frac{\sqrt{2}}{v_2} m_f(T = 0)\,.
\end{equation}

\subsection{Scalar masses}\label{s:app:2hdmmasses:scalars}
The scalar mass matrix is obtained from the second derivative of the tree-level potential $V_{\text{tree}}$ of Eq.~\eqref{eq:v2hdm}. When all $\bar\omega_i$ are non-zero, the scalar mass matrix is a $8\times8$ matrix. In the case of a neutral CP-conserving vacuum, where only $\bar\omega_{1, 2}$ are non-zero, the mass matrix becomes block-diagonal with a different block for the charged, the neutral CP-even, and the neutral CP-odd field components. These blocks are always symmetric $2\times 2$ matrices of the form
\begin{equation}
	\mathcal{M} =
		\begin{pmatrix}
			\mathcal{M}_{11} & \mathcal{M}_{12}\\
			\mathcal{M}_{12} & \mathcal{M}_{22}
		\end{pmatrix},
\end{equation}
with the two solutions
\begin{equation}\label{eq:masseigenvalues}
	m^2 = \frac12\left(\mathcal{M}_{11} + \mathcal{M}_{22}\right) \pm \frac12\sqrt{\left(\mathcal{M}_{11} - \mathcal{M}_{22}\right)^2 + 4\mathcal{M}_{12}^2}\,.
\end{equation}
The thermal corrections from the resummation of the daisy diagrams can then be included via
\begin{equation}
	\overline{\mathcal{M}} = \mathcal{M} + \operatorname{diag}\left(c_1, c_2\right) T^2 =
		\begin{pmatrix}
			\mathcal{M}_{11} + c_1 T^2 & \mathcal{M}_{12}\\
			\mathcal{M}_{12} & \mathcal{M}_{22} + c_2 T^2
		\end{pmatrix},
\end{equation}
with the thermal coefficients $c_{1,2}$ given in Eqs.~\eqref{eq:c1coeff} and \eqref{eq:c2coeff}, leading to the masses including thermal corrections
\begin{equation}
	\overline m^2 = \frac12\left(\mathcal{M}_{11} + \mathcal{M}_{22} + \left(c_1 + c_2\right) T^2\right) \pm \frac12\sqrt{\left(\mathcal{M}_{11} - \mathcal{M}_{22} + \left(c_1 - c_2\right) T^2\right)^2 + 4\mathcal{M}_{12}^2}\,.
\end{equation}

\subsubsection{Charged sector}
For the charged sector $C$, the matrix elements are given by
\begin{align}
	\mathcal{M}^C_{11} &= m_{11}^2 + \frac{\lambda_1}{2}\bar\omega_1^2 + \frac{\lambda_3}{2}\bar\omega_2^2\,,\\
	\mathcal{M}^C_{22} &= m_{22}^2 + \frac{\lambda_2}{2}\bar\omega_2^2 + \frac{\lambda_3}{2}\bar\omega_1^2\,,\\
	\mathcal{M}^C_{12} &= -m_{12}^2 + \frac{\lambda_4 + \lambda_5}{2}\bar\omega_1 \bar\omega_2\,.
\end{align}
This matrix is diagonalised by the angle $\beta$ of Eq.~\eqref{eq:tanbeta}, and its eigenvalues can be written for the EW vacuum at $T = 0$ in the usual form,
\begin{align}
    m_{G^\pm} &= 0\,,\\
    m_{H^\pm} &= \left[\frac{m_{12}^2}{v_1 v_2} - \frac12\left(\lambda_4 + \lambda_5\right)\right] v^2\,,\label{eq:chargedhiggsmass}
\end{align}
where $G^\pm$ denotes the massless charged Goldstone boson, and $H^\pm$ the charged Higgs boson. Note that we use the Landau gauge.

\subsubsection{Neutral CP-odd sector}
For the neutral CP-odd sector $P$, the matrix elements are
\begin{align}
	\mathcal{M}^P_{11} &= m_{11}^2 + \frac{\lambda_1}{2}\bar\omega_1^2 + \frac{\lambda_3 + \lambda_4 - \lambda_5}{2}\bar\omega_2^2\,,\\
	\mathcal{M}^P_{22} &= m_{22}^2 + \frac{\lambda_2}{2}\bar\omega_2^2 + \frac{\lambda_3 + \lambda_4 - \lambda_5}{2}\bar\omega_1^2\,,\\
	\mathcal{M}^P_{12} &= -m_{12}^2 + \lambda_5\bar\omega_1 \bar\omega_2\,.
\end{align}
Also this matrix is diagonalised by the angle $\beta$ of Eq.~\eqref{eq:tanbeta}, and we get the two eigenvalues at $T = 0$,
\begin{align}
    m_{G^0} &= 0\,,\\
    m_A &= \left(\frac{m_{12}^2}{v_1 v_2} - \lambda_5\right) v^2\,,\label{eq:cpoddhiggsmass}
\end{align}
where $G^0$ denotes the massless neutral Goldstone boson, and $A$ the CP-odd Higgs boson.

\subsubsection{Neutral CP-even sector}
Finally, for the neutral CP-even sector $S$, we have
\begin{align}
	\mathcal{M}^S_{11} &= m_{11}^2 + \frac{3\lambda_1}{2}\bar\omega_1^2 + \frac{\lambda_3 + \lambda_4 + \lambda_5}{2}\bar\omega_2^2\,,\\
	\mathcal{M}^S_{22} &= m_{22}^2 + \frac{3\lambda_2}{2}\bar\omega_2^2 + \frac{\lambda_3 + \lambda_4 + \lambda_5}{2}\bar\omega_1^2\,,\\
	\mathcal{M}^S_{12} &= -m_{12}^2 + \left(\lambda_3 + \lambda_4 + \lambda_5\right)\bar\omega_1 \bar\omega_2\,.
\end{align}
The rotation matrix diagonalising this matrix can be written in terms of a mixing angle $\alpha$. Due to the two eigenvalues $m_{h,H}$ being relatively long analytical expressions, which are given by the two solutions in Eq.~\eqref{eq:masseigenvalues}, we refrain from writing them out explicitly and only state that by convention $m_{h(H)}$ corresponds to the lighter (heavier) CP-even Higgs boson. We choose our parameters such that $h$ corresponds to the SM-like Higgs boson with mass 125~GeV.

\subsection{Gauge boson masses}\label{s:app:2hdmmasses:gaugebosons}
The mass matrix for the gauge bosons in the interaction basis $(W^1, W^2, W^3, B^0)$ is given by
\begin{equation}
	\mathcal{M}_{\mathrm{g}} =
		\begin{pmatrix}
			\frac14 g^2 \bar\omega^2 & 0 & 0 & 0 \\
			0 & \frac14 g^2 \bar\omega^2 & 0 & 0 \\
			0 & 0 & \frac14 g^2 \bar\omega^2 & -\frac14 g g' \bar\omega^2 \\
			0 & 0 & -\frac14 g g' \bar\omega^2 & \frac14 g'^2 \bar\omega^2
		\end{pmatrix}\,,
\end{equation}
with the $SU(2)$ and $U(1)$ gauge couplings $g$ and $g'$, respectively, and $\bar\omega^2 \equiv \bar\omega_1^2 + \bar\omega_2^2$. Its diagonalisation leads to the well-known eigenvalues at $T = 0$ for the $W^\pm$, $\gamma$, and $Z$ bosons,
\begin{equation}
	m_W^2 = \frac{g^2}{4} v^2\,,\qquad	m_\gamma^2 = 0\,,\qquad	m_Z^2 = \frac{g^2 + g'^2}{4} v^2\,.
\end{equation}
Similarly to the scalar sector, the thermal corrections can be included by adding an additional term to the mass matrix, and in theories with only extended Higgs sectors (i.e.\ no additional fermions, coloured scalars, or vector bosons), this term is diagonal, and, with the exception of the gauge couplings $g$ and $g'$, universal for each component. Note that only the longitudinal degrees of freedom of the gauge bosons will receive thermal corrections, leading also to a finite longitudinal photon mass. Labelling the thermal coefficient $c_{\mathrm{g}}$, we have:
\begin{equation}
	\overline{\mathcal{M}}_{\mathrm{g}} = \mathcal{M}_{\mathrm{g}} + c_{\mathrm{g}} \operatorname{diag}\left(g^2, g^2, g^2, g'^2\right) T^2\,,
\end{equation}
with \cite{Basler:2018cwe}
\begin{equation}
	c_{\mathrm{g}} = \frac13\left(\frac{n_H}{8} + 5\right) =
		\begin{cases}
			\frac{11}{6} & \text{ for the SM with } n_H = 4\,,\\
			2 & \text{ for the 2HDM with } n_H = 8\,,
		\end{cases}
\end{equation}
where $n_H$ denotes the number of Higgs and Goldstone fields coupling to the gauge bosons. Then, the thermally corrected longitudinal gauge boson masses become,
\begin{align}
	\overline m_W^2 &= \frac{g^2}{4}\bar\omega^2 + c_{\mathrm{g}} g^2 T^2\,,\\
	\overline m_\gamma^2 &= \frac{g^2 + g'^2}{2}\left(\frac{\bar\omega^2}{4} + c_{\mathrm{g}} T^2\right) - \Delta\,,\\
	\overline m_Z^2 &= \frac{g^2 + g'^2}{2}\left(\frac{\bar\omega^2}{4} + c_{\mathrm{g}} T^2\right) + \Delta\,,
\end{align}
with
\begin{equation}
	\Delta = \sqrt{\frac{\left(g^2 + g'^2\right)^2}{4}\left(\frac{\bar\omega^2}{4} + c_{\mathrm{g}} T^2\right)^2 - g^2 g'^2\left(\frac{\bar\omega^2}{2} + c_{\mathrm{g}} T^2\right) c_{\mathrm{g}} T^2}\,.
\end{equation}

\section{Potential parameter dependences}\label{s:app:lambdas}
\begin{figure}[tp]
	\centering
	\includegraphics[width=\textwidth]{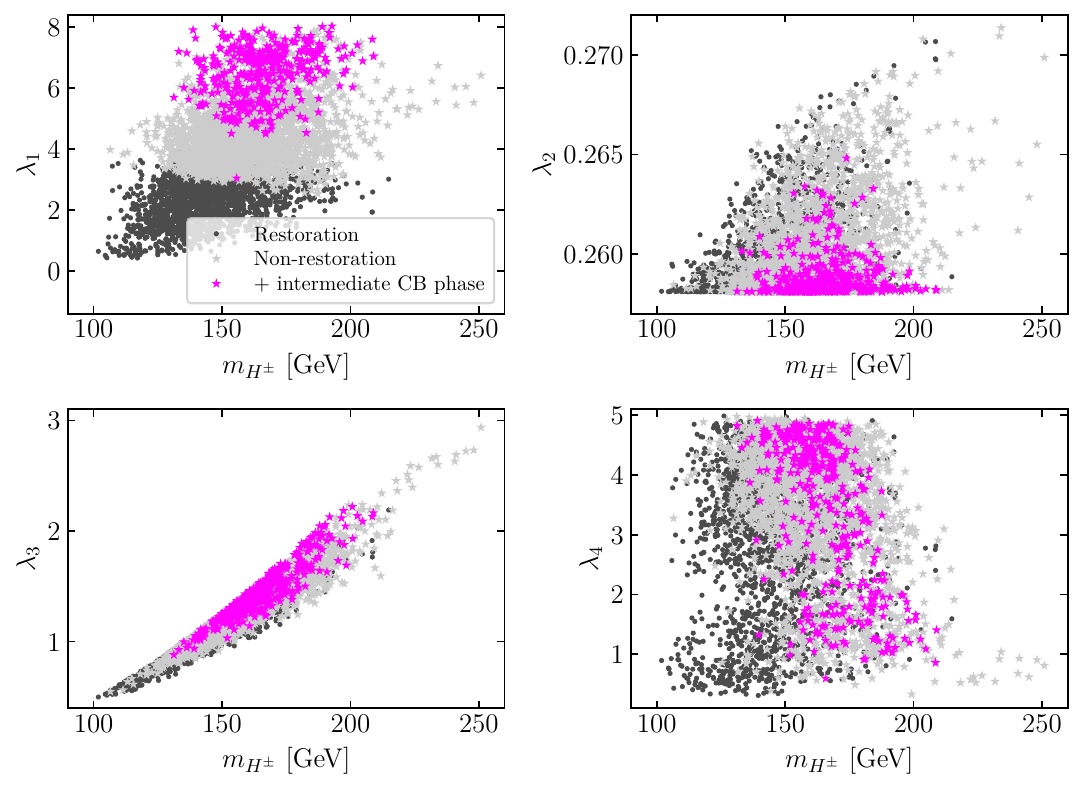} \\
	\includegraphics[width=.55\textwidth]{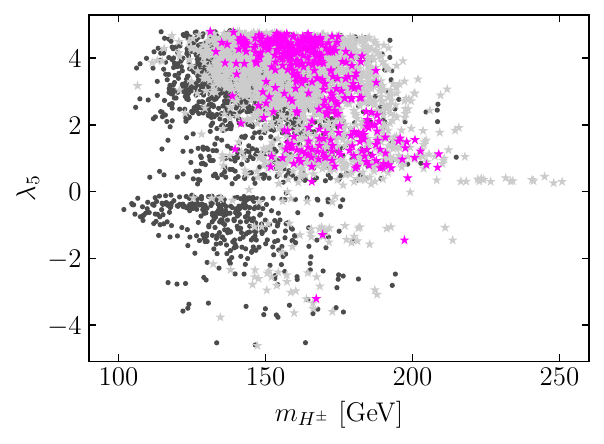}
	\caption{Parameter points of our scan shown for $\lambda_i$ ($i=1,...,5)$ as a function of $m_{H^\pm}$. All points fulfill the experimental constraints from \ScannerS{}. Dark (light) grey points: EW symmetry (non-)restoration at high temperatures. Magenta points exhibit in addition an intermediate CB phase at non-zero temperature using the one-loop effective potential.}
	\label{fig:scan2lambdam12}
\end{figure}
\begin{figure}[tp]
	\centering
	\includegraphics[width=.5\textwidth]{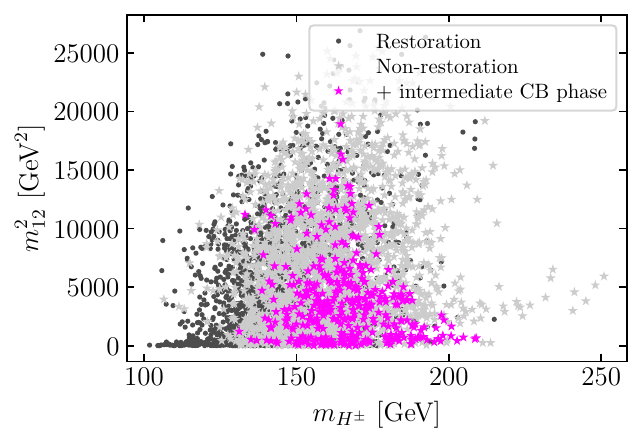}\includegraphics[width=.5\textwidth]{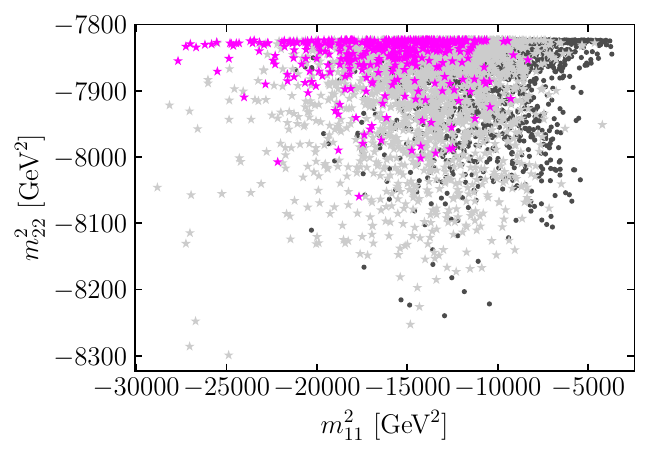}
	\caption{Parameter points of our scan shown for $m_{12}^2$ as a function of $m_{H^\pm}$ (left), as well as $m_{22}^2$ versus $m_{11}^2$ (right). All points fulfill the experimental constraints from \ScannerS{}. Dark (light) grey points: EW symmetry (non-)restoration at high temperatures. Magenta points exhibit in addition an intermediate CB phase at non-zero temperature using the one-loop effective potential.}
	\label{fig:scan2m11m12m22}
\end{figure}
Figures~\ref{fig:scan2lambdam12} and \ref{fig:scan2m11m12m22} show the distributions of the potential parameters $\lambda_i$ ($i=1,...,5$) and $m_{12}^2$, respectively, as a function of the charged Higgs mass, as well as $m_{11}^2$ versus $m_{22}^2$ for all points of our scan compatible with the relevant theoretical and experimental constraints. Dark (light) grey points (do not) fulfill symmetry restoration at high temperatures. Magenta points exhibit in addition an intermediate CB phase at non-zero temperature obtained from the minimization of the one-loop effective potential.